\documentclass[a4paper,11pt]{article}
\pdfoutput=1 

\usepackage{jheppub} 

\usepackage[T1]{fontenc} 
\usepackage{amsmath}
\usepackage{amsfonts}
\usepackage{amsmath}
\usepackage{amsxtra}
\usepackage{amstext}
\usepackage{amssymb}
\usepackage{slashed}
\usepackage{graphicx}
\usepackage{cleveref}
\usepackage{color}

\numberwithin{equation}{section}

\newcommand{\be}{\begin{equation}}
\newcommand{\ee}{\end{equation}}
\newcommand{\bea}{\begin{eqnarray}}
\newcommand{\eea}{\end{eqnarray}}

\newcommand{\cN}{\mathcal{N}}
\newcommand{\cB}{\mathcal{B}}
\newcommand{\jp}{j_+}
\newcommand{\xe}{x_F}

\newcommand{\cO}{\mathcal{O}}

\def\d{\mathrm{d}}

\def\({\left(}
\def\){\right)}
\def\[{\left[}
\def\]{\right]}
\newcommand{\rvline}{\hspace*{-\arraycolsep}\vline\hspace*{-\arraycolsep}}

\def\-{\,-\,}
\def\={\,=\,}
\def\+{\,+\,}
\def\equi{\,\equiv\,}

\newcommand{\adstwo}{\ensuremath{\mathrm{AdS}_2}}

\newcommand{\stwo}{\ensuremath{\mathrm{S}^2}}
\newcommand{\sthree}{\ensuremath{\mathrm{S}^3}}

\definecolor{cardinal}{rgb}{0.6,0,0}
\definecolor{darkgreen}{rgb}{0,0.4,0}
\definecolor{purple}{rgb}{0.5, 0, 0.5}
\definecolor{golden}{rgb}{0.92, 0.7, 0}
\definecolor{midnight}{rgb}{0, 0, 0.5}
\definecolor{darkblue}{rgb}{0, 0, 0.7}

\title{\boldmath Black Holes and the Swampland: the Deep Throat revelations}

\author[a]{Yixuan Li }

\affiliation[a]{Institut de Physique Théorique, Université Paris-Saclay, CNRS, CEA, \\
91191, Gif-sur-Yvette, France.}

\emailAdd{yixuan.li@ipht.fr}

\abstract{Multi-centered bubbling solutions are black hole microstate geometries that arise as smooth solutions of 5-dimensional $\cN=2$ Supergravity. When these solutions reach the scaling limit, their resulting geometries develop an infinitely deep throat and look arbitrarily close to a black hole geometry. We depict a connection between the scaling limit in the moduli space of Microstate Geometries and the Swampland Distance Conjecture. The naive extension of the Distance Conjecture implies that the distance in moduli space between a reference point and a point approaching the scaling limit is set by the proper length of the throat as it approaches the scaling limit. Independently, we also compute a distance in the moduli space of 3-centre solutions, from the K\"ahler structure of its phase space using quiver quantum mechanics. We show that the two computations of the distance in moduli space do not agree and comment on the physical implications of this mismatch.
}

\begin{document} 
\maketitle
\flushbottom

\section{Introduction}
\label{sec:intro}

The fact that black holes are statistical objects with temperature and entropy raises two key issues. First, how to describe the microstates accounting for the statistical entropy? Second, how does the black hole restore the information that falls in it?
String Theory's historical answer to the first question is to describe the microstates at low string coupling, where all the possible open strings that stretch between brane bound states have an entropy that matches the statistical entropy of the black hole.
As the string coupling constant is tuned to larger values to a regime where gravity is dominant, the branes expand in size, so one could expect that the microstates differ from the black hole at horizon-scale.

The Fuzzball paradigm \cite{Mathur:2005zp,Skenderis:2008qn} proposes that black hole microstates do possess a horizon-size structure that differs from the classical black hole. Within this approach, it is expected that the black hole evaporation is similar to the burning of a star or a piece of coal \cite{Avery:2009tu}, differing from Hawking's calculation which leads to the Information Paradox.
%
Within the Fuzzball paradigm, the Microstate Geometries programme \cite{Bena:2007kg,Warner:2019jll} endeavours to describe these black hole microstates within the Supergravity approximation of String Theory by smooth horizonless solutions. If one succeeds in finding a large number --- hopefully $e^S$ --- of them, then one has answered to the question ``What do black hole microstates look like?''.

\begin{figure}
\includegraphics[width=0.5\linewidth]{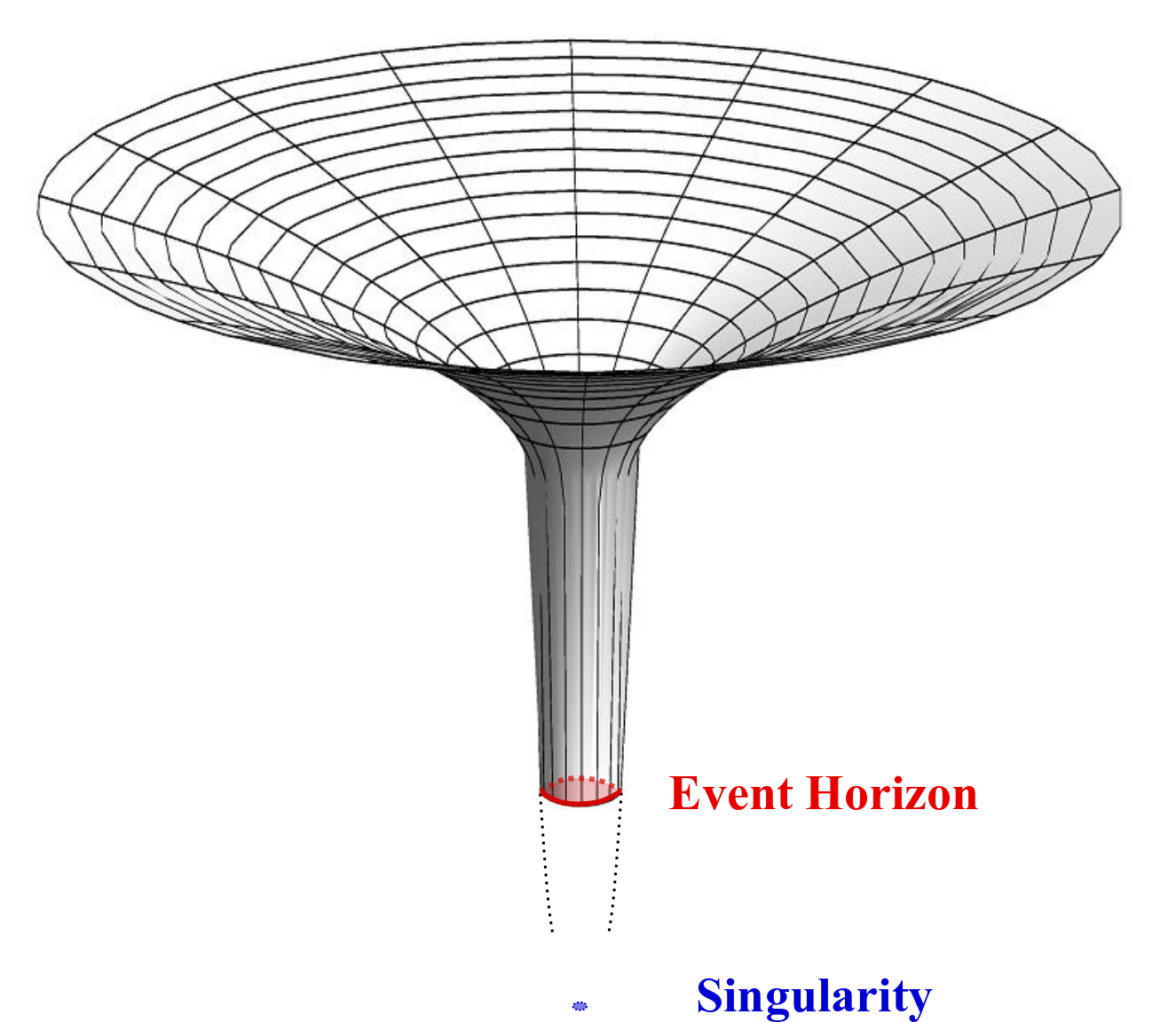}
\includegraphics[width=0.5\linewidth]{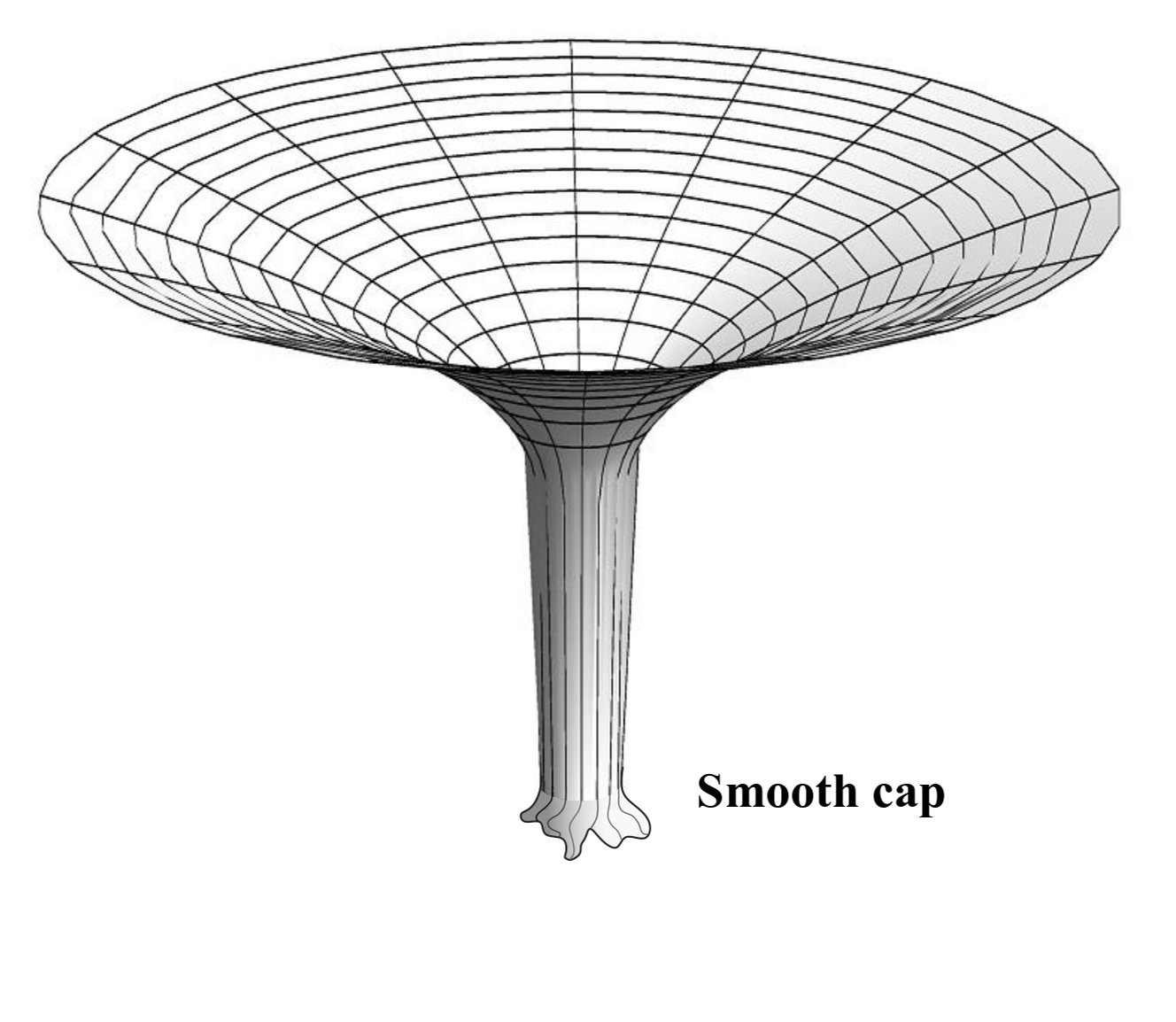}
\caption{Schematic picture of a classical black hole (left) and a microstate geometry (right). From the asymptotic observer's point of view, instead of the event horizon lying at the bottom of a throat of infinite length, a microstate geometry would replace it with a smooth cap lying at the bottom of a throat of finite length. In the scaling limit, its throat length increases to infinity while the cap's geometry stays constant. The figures are from \cite{Heidmann:2019gvg}.}
\label{fig:FuzzballProp}
\end{figure}

In many classes of microstate geometries, the infinitely-long throat of an extremal black hole is replaced by a cap at the end of a long, but finite throat \cite{Bena:2006kb,Bena:2015bea,Bena:2016ypk} (See Fig. \ref{fig:FuzzballProp}). The procedure to construct a large number of Supergravity microstates is the following: Take a black hole with given charges and angular momenta. Supergravity admits a large number of solutions with finite throat length, with charges and angular momenta equal to those of the black hole. In the moduli space (of a particular superselection sector, if any),\footnote{In some models of microstate geometries, as in Multi-centered bubbling models, there are families of solutions labeled by the fluxes $\Gamma_i$ wrapping the bubbles (see Section \ref{sec:2}). Inside each of these families, or \textit{superselection sectors} \cite{Bena:2013gma}, there are still real parameters left to characterize the solutions, defining a moduli space. There are restrictions on the bubble fluxes (and on the superselection sectors) to admit a scaling limit; but here, we consider one superselection sector which does.}
each of these solutions admits a limit --- called the \textit{scaling limit} --- where, from the perspective of an observer at infinity, they become more and more similar to the black hole; in particular, their throat length increases to infinity in the scaling limit, while the size of the cap remains fixed.\footnote{Infinitely-long throats arise in extremal black holes, but not in non-extremal ones, so this procedure to construct microstate geometries \textit{a priori} applies only for extremal black holes. Although in this paper we will only consider a class of extremal BPS black holes and their microstate geometries, scaling solutions can arise in similar non-BPS extremal black holes as well \cite{Bena:2009ev,Bena:2009en,Bah:2021jno}.}

In the moduli space of solutions, the scaling limit point plays a particular role, for the following reasons: \newline
(1) The scaling limit lies at the boundary of moduli space where the throat length increases to infinity. \newline
(2) As we approach the scaling limit, global symmetries of the black hole, which obeys the no-hair theorem, are restored. For instance, microstate geometries do not generically possess the $\mathrm{SO}(3)$-rotational symmetry of the black hole. \newline
(3) Everywhere in moduli space, energy excitations at the bottom of the throat of deep microstates are gapped. Their gap matches that of their dual CFT states \cite{Bena:2006kb,Bena:2018bbd}. However, in the scaling limit, the mass gaps of these modes decrease to zero, because of the increasing redshift due to the lengthening of the throat.

Taking the limit to a boundary point of moduli space is reminiscent of the Swampland Distance Conjecture \cite{Ooguri:2006in}, that we are reformulating hereinbelow. Consider an effective field theory consistent with quantum gravity, with an arcwise-connected moduli space --- a point in the moduli space fixes the expectation value of the scalar fields of the EFT. The Swampland Distance Conjecture states that: \newline
(1) The moduli space is not bounded in terms of its geodesic distance $d$. In other words, given $p_0$ a point in the bulk of the moduli space, there exists a family of arcwise-connected points $\{p\}$ going from $p_0$ to an infinite geodesic distance with respect to $p_0$. \newline
(2) Global symmetries are restored at infinite distance in moduli space. \cite{Grimm:2018ohb} \newline
(3) Given the point $p_0$ and the path of points $\{p\}$ defined in (1), there exists $\alpha>0$ and there exists an infinite tower of states with an associated mass scale $M(p)$ such that 
\be \label{eq:Swampland_distance_conjecture}
M(p) \underset{d(p_0,p)\rightarrow \infty}{\sim} M(p_0) \, e^{-\alpha d(p_0,p)} \,.
\ee

Originally, the distance in moduli space was defined according to the kinetic terms of the scalar fields in the EFT in the following sense: Consider a $d$-dimensional EFT whose action in the $d$-dimensional Einstein frame is written as
\be
S = \int d^d x \sqrt{-g} \left[\frac{R}{2} - g_{ij}\left( \phi^i \right) \partial \phi^i \partial \phi^j + ... \right] \;.
\ee
Then $g_{ij}$ defines a metric on the moduli space of effective field theories.
Following \cite{Lust:2019zwm}, it has been proposed to generalise the Swampland Distance Conjecture --- about moduli spaces of scalar fields --- to a space of metrics.
A notion of distance can be defined on a transverse-traceless metric $g_{\mu\nu}$ of a spacetime $M$ of volume $V_{M}=\int_{M} \sqrt{g}$ \cite{DeWitt:1967yk}
\be \label{eq:generalized_distance}
\Delta_\textrm{generalized}=c \int_{\tau_{i}}^{\tau_{f}}\left(\frac{1}{V_{M}} \int_{M} \sqrt{g} \operatorname{tr}\left[\left(g^{-1} \frac{\partial g}{\partial \tau}\right)^{2}\right]\right)^{\frac{1}{2}} \mathrm{d} \tau \,.
\ee
This distance boils down to the moduli space of the scalar fields in the case of Calabi-Yau compactifications on 4-dimensional Minkowski space \cite{Candelas:1990pi}.

Thanks to this notion of distance between two metrics, it was argued in \cite{Lust:2019zwm} that the vanishing limit of the negative cosmological constant, $\Lambda$, in an AdS vacuum in String Theory leads to an infinite tower of light states --- for instance the tower of Kaluza-Klein modes of some decompactifying parts from the internal manifold. Using the distance (\ref{eq:generalized_distance}), the authors of \cite{Bonnefoy:2019nzv} computed distances on the space of black holes metrics and related the infinite black-hole-entropy limits to both massless Kaluza-Klein modes of an internal Calabi-Yau manifold and possibly Goldstone modes of BMS-like transformations on the black hole horizon.
\newline

To extend the Swampland Distance Conjecture for metrics to the scaling limit of microstate geometries, we would like to show that the infinite tower of gapped modes on top of any microstate geometry collapses, and the masses of all these modes decrease exponentially. 
In this paper, we study a class of \textit{bubbling microstate geometries}, that descend in 4 dimensions to \textit{multicentered solutions} \cite{Denef:2000nb,Bates:2003vx}. In the scaling limit, different microstate geometries approach the BMPV black hole \cite{Breckenridge:1996is}. These microstate geometries possess an $\adstwo \times \sthree$ throat. 
We will show that, in the scaling limit, the mass the Kaluza-Klein modes of the $\sthree$ measured by an observer at spacial infinity decrease exponentially with respect to the length of the throat. At the bottom of this throat lie also non-trivial two-cycles; the mass of the M2 branes wrapping these two-cycles decrease exponentially in the same fashion in the scaling limit.
By reading off the expression inside the exponential, one can infer the distance in moduli space $\Delta_\textrm{exponential}$ that would be in agreement with the extension of the Swampland Distance Conjecture; one thus expects this distance to be proportional to the length of the $\adstwo$ throat, which is becoming \textit{infinite} in the scaling limit. 
As a result, our study quite possibly extends the Swampland Distance Conjecture in a rather unusual way.

In addition, we will also compare this distance with another notion of distance in the moduli space of solutions, whose computation is independent of Swampland notions. Out of a Lagrangian theory characterizing a set of fields $\phi^A$, the symplectic form, $\Omega$, of the theory can be defined, from the Crnkovi\'c-Witten-Zuckerman formalism, as an integral over a Cauchy surface $\Sigma$ \cite{Witten:1986qs,Crnkovic:1987tz} 
\be
\Omega = \int d\Sigma_l \,\, \delta\left( \frac{\partial L}{\partial (\partial_l \phi^A)}\right) \wedge \delta \phi^A\, .
\ee
If $\Omega$ is closed and non-degenerate, the $2m$-dimensional solution-space manifold is reinterpreted as the \textit{phase space}, whose symplectic volume (in units of $h^{m}$) gives the number of microscopic ground states.
When the phase-space manifold is furthermore endowed with an integrable complex structure, $J$, and if $\Omega(\cdot, J\cdot)$ is a Riemannian metric, the manifold is K\"ahler and one can define a distance, $\Delta_\mathrm{phase}$, on the moduli space of solutions using the K\"ahler metric $\Omega(\cdot, J\cdot)$.
Luckily, the solution space of \textit{three-centered multicenter solutions} one constructs as Microstate geometries is a K\"ahler manifold \cite{deBoer:2008zn}; so we will measure the distance to the scaling limit with respect to this K\"ahler metric. 

Surprisingly, we find that with respect the ``canonical'' $\Delta_\mathrm{phase}$ that would be in agreement with computations in \cite{deBoer:2008zn}, the scaling limit lies at \textit{finite} distance in moduli space, in tension with the distance $\Delta_\mathrm{exponential}$.
However, the computation of $\Delta_\mathrm{phase}$ is performed at weak string coupling regime using quiver quantum mechanics, and one can wonder whether this computation is still reliable in the regime where Supergravity dominates. However, in \cite{deBoer:2008zn}, the authors argue that the reduced symplectic form does not vary with the string coupling constant thanks to a non-renormalization theorem, and further conclude that Supergravity is breaking down because of large quantum fluctuations in scaling geometries, and hence could not be a good description of these geometries.
From the weak-coupling regime, they also infer that the scaling limit, which was perfectly in reach within Supergravity, is actually prohibited if one accounts for quantum effects, which prevent the quantum wave functions to populate the region of classical moduli space close to the scaling limit.
Thus, if the correct normalization of distance on moduli space is given by $\Delta_\mathrm{exponential}$ and not by $\Delta_\mathrm{phase}$, then the breakdown of Supergravity due to quantum effects prescribed in \cite{deBoer:2008zn} would be softened.
\newline

The organisation of the paper is the following. In Section \ref{sec:2}, we review smooth multi-centered bubbling solutions in five dimensions (and their M-theory uplift) of \cite{Bena:2006kb}.
In Section \ref{sec:3}, we first compute how the throat lengths of bubbling solutions behave in the scaling limit. We then find an exponential decrease of the $\sthree$ Kaluza-Klein mass tower, consistent with a naive extension of the Swampland Distance Hypothesis to this system.
In Section \ref{sec:4}, we study the moduli space of three-centre bubbling solutions, independently of the Swampland programme. Using the results of \cite{deBoer:2008zn}, we determine the metric on moduli space of solutions coming from the symplectic form.
We show that with this distance, $\Delta_\mathrm{phase}$, the scaling limit lies at finite distance in moduli space  and that all the moduli space is bounded.
In section \ref{sec:5}, we discuss the tension between the two distances on moduli space and share some insight about the ability of Supergravity to describe black hole microstates with arbitrarily deep throats.

\section{Multicenter bubbling solutions}
\label{sec:2}

\subsection{Multicenter bubbling solutions in 5 and 11 dimensions}

Upon compactifying maximal eleven-dimensional Supergravity on Calabi-Yau threefold, the resulting five-dimensional $\cN=2$ supergravity coupled to $n_V$ vector multiplets with $n_V \leq 2$ contains the following bosonic fields:
\begin{itemize}
    \item a gravitational field, $g$, 
    \item $n_V+1$ U(1) vector gauge fields, $A^{I}_\mu$, whose field strengths are denoted $F^I= d_5 A^I$,
    \item $n_V+1$ scalars, $X^I$.
\end{itemize}
This theory is described by the action
\be 
(16\pi G_{5})\,S_{5} = \int d^{5}x\, \sqrt{-g}\, R \- Q_{IJ}  \int \,\left(F^I\wedge \star_{5} F^J - d_5 X^I \wedge \star_{5} d_5X^J\right) \+ \frac{C_{IJK}}{6}\,\int A^I \wedge F^J\wedge F^K\,,
\label{eq:c25dAction}
\ee
where $C_{IJK}$ are the structure constants satisfying the fixed-volume constraint
\be
\frac{1}{6} \, C_{IJK}\, X^I X^J X^K \= 1 \qquad \Longrightarrow \qquad X_I \=  \frac{1}{6}\,C_{IJK}\, X^J X^K\,,
\label{eq:c2FixedVolCompact}
\ee
and the couplings $Q_{IJ}$ depend on the scalars via
\be 
Q_{IJ} \= \frac{9}{2}\, X_I X_J -\frac{1}{2} \, C_{IJK} X^K\, .
\ee

The action admits the following Einstein-Maxwell-scalar equations of motion
\be 
\begin{split}
R_{\mu \nu} \+ Q_{IJ} \left(\partial_\mu X^I \partial_\nu X^J \+ F^I_{\mu \rho}\, {F_{\nu}^J}^{ \rho} \- \frac{1}{6}\, g_{\mu \nu}\, F^I_{ \rho \sigma}\, {F^J}^{ \rho \sigma} \right) &\= 0\, ,\\
d_{5}\left(Q_{IJ}\,\, \star_{5} F^J \right)\+ \frac{1}{4}\,C_{IJK}\, F^J \wedge F^K &\= 0\,, \\
-d_5 \star_5 d_5 X_I \+ \left( C_{IJK} X_L X^K - \frac{1}{6} C_{ILJ}\right) \left( F^L\wedge \star_{5} F^J - dX^L \wedge \star_{5} dX^J \right) &\= 0\, .
\end{split}
\label{eq:c3EinsteinMaxScal5d}
\ee
\newline


The most general supersymmetric solution to $\cN=2$ five-dimensional Supergravity coupled to $n_V$ extra gauge fields with structure constant $C_{IJK}$, admitting a time-like Killing vector $\partial_t$ are characterized by $n_V+1$ electric warp factors $Z_I$, $n_V+1$ magnetic self-dual two-forms $\Theta^I$, an angular momentum one-form $\omega$, and a space-like hyper-K\"ahler manifold $\cB$. The metric and the field strengths are stationary, and are split in the following way \cite{Gutowski:2004yv,Bena:2004de}:
\be
\begin{split}
ds_5^2  & \= -\left(\frac{1}{6} C_{IJK} \,Z_I Z_J Z_K \right)^{-\frac{2}{3}} \left(dt+\omega \right)^2 \+ \left(\frac{1}{6} C_{IJK} \,Z_I Z_J Z_K \right)^{\frac{1}{3}} \, ds\left(\cB\right)^2 \,,\\
F^I &\= d_4 A^I \= d_4 \left( Z_I^{-1} \left(dt+\omega \right) \right) \+ \Theta^I\,.
\end{split}
\label{eq:c35dmetric&fields}
\ee
In terms of these new data, the Einstein-Maxwell-scalar equations of motion (\ref{eq:c3EinsteinMaxScal5d}) are rewritten as the so-called BPS equations
\begin{align}
        \star_4 \Theta^I &\= \Theta^I \, , \quad \textrm{with  } d_4 \Theta^I \= 0 \, ,  \label{eq:1st_BPS_eq}\\
    {\nabla_4}^2V \equiv \star_4 d_4 \star_4 d_4 \, Z_I &\= \frac{1}{2} C_{IJK} \, \star_4 \left( \Theta^J \wedge \Theta^K \right) \,, \label{eq:2nd_BPS_eq}\\
    d_4 \omega \+ \star_4 d_4 \omega &\= Z_I \, \Theta^I\,. \label{eq:3rd_BPS_eq}
\end{align}
The first set of $n_V+1$ equations (\ref{eq:1st_BPS_eq}) ($I = 1,\ldots, n_V+1$) determine the magnetic two-forms.
The second set of $n_V+1$ equations (\ref{eq:2nd_BPS_eq}) determine the electric warp factors, sourced by the magnetic fields. The fact that magnetic fluxes source a net electric charge is made possible thanks to the Cherns-Simons term in the five-dimensional Supergravity action (\ref{eq:c25dAction}); this is essential in the construction of smooth solitonic solutions in Supergravity.
The last equation (\ref{eq:3rd_BPS_eq}) tells that the angular momentum $\omega$ is sourced by electric and magnetic fields, recalling the Poynting vector in electromagnetism.
\newline

We now consider $\cB$ to be a four-dimensional Gibbons-Hawking space. The Gibbons-Hawking space is made of multiple centers of Kaluza-Klein monopoles. 
The Gibbons-Hawking space possesses non-trivial two-cycles called bubbles, defined by the shrinking of the coordinate $\psi$ fibered along any line running between a pair of Gibbons-Hawking points in $\mathbb{R}^3$. The spatial part of the metric in (\ref{eq:c35dmetric&fields}) is thus an $\mathrm{S}^1$ fibered along $\mathbb{R}^3$; it is determined by a harmonic function $V$ in $\mathbb{R}^3$ (${\nabla_3}^2V \equiv \star_3 d_3 \star_3 d_3 V\= 0$) and a one-form $A$ (with $\nabla_3 A \equiv \star_3 d_3 A = d_3 V $):
\be 
ds\left(\cB\right)^2  \= V^{-1} \, \left( d\psi +A \right)^2 \+ V  \,\left[\,d\rho^2 + \rho^2 \, \left( d\vartheta^2 + \sin^2 \vartheta \, d\phi^2 \right)\,\right]\,.
\label{eq:c3GHmetric}
\ee
The potential $V$ is sourced by a set of $n$ Gibbons-Hawking centres labeled by $j$, of charge $q_j$:
\be 
V(\vec{\rho}) \= h_\infty \+ \sum_{j=1}^n \frac{q_j}{\rho_j}\, , \qquad A \= \sum_{j=1}^n q_j \,\cos\vartheta_j \,d\phi_j \,,
\label{eq:c3GHharmonicfunction}
\ee
where ($\rho_j,\vartheta_j,\phi_j$) are the shifted spherical coordinates around the $j^\text{th}$ center. The potential $V$ is a harmonic function on $\mathbb{R}^3$. The Gibbons-Hawking space pinches off smoothly around each center $j$: the geometry is a flat $\mathbb{R}^4$ modded by $\mathbb{Z}_{|q_j|}$ along $\psi$, where $q_j \in \mathbb{Z}$.
Besides, $\mathbb{R}^4$ is asymptotically modded by $\mathbb{Z}_{\sum |q_j|}$, so it is convenient to subsequently impose $\sum_j |q_j|=1$ to have an asymptotic $\mathbb{R}^4$.

We will consider solutions that are independent of $\psi$. With this assumption, the other solution data --- $Z_I$, $\Theta^i$ and $\omega$ --- are all given in terms of harmonic functions on $\mathbb{R}^3$.

The $n_V+1$ self-dual magnetic two-forms $\Theta^I$ are of the form
\be 
\Theta^I \= \partial_a \left( V^{-1} \, K^I \right) \Omega^a \, ,
\ee
where $(\Omega^1,\Omega^2,\Omega^3)$ is a basis of self-dual (in 4 dimensions) two-forms and $K^I$ are harmonic functions on $\mathbb{R}^3$ of the form
\be 
K^I \= k^I_\infty \+ \sum_{j=1}^n \frac{k^I_j}{\rho_j} \, .
\ee
The number $k^I_i-k^i_j$ is the magnetic flux on the two-cycle between centres $i$ and $j$. 

The $n_V+1$ warp factors $Z_I$ are
\be 
Z_I \= L_I \+ \frac{C_{IJK}}{2} \frac{K^J K^K}{V}
\label{eq:Z_I_intermsof_harmonic}
\ee
where $L_I$ is a harmonic function on $\mathbb{R}^3$ is
\be 
L_I \= l^I_\infty \+ \sum_{j=1}^n \frac{l^I_j}{\rho_j} \, .
\label{eq:c3LIfunctions}
\ee
From the 5-dimensional Supergravity perspective, $l^I_j$ is the electric charge of $L_I$ at the $j^\text{th}$ center. 


Finally, the angular-momentum one-form can be decomposed along the U(1) $\psi$-fiber:
\be 
\omega = \left( M \+ \frac{K^I L_I }{2\,V}\+ \frac{C_{IJK}}{6} \frac{K^I K^J K^K}{V^2} \right) \left( d\psi + A\right) \+ \varpi 
\equiv \mu \left( d\psi + A\right) \+ \varpi \,,
\ee
where $\varpi$ is a one-form on $\mathbb{R}^3$ and $M$ --- the harmonic conjugate of $\omega$ in (\ref{eq:3rd_BPS_eq}) --- is a harmonic function on $\mathbb{R}^3$ of the form
\be 
M \= m_\infty \+ \sum_{j=1}^n \frac{m_j}{\rho_j} \,.
\label{eq:c3Mfunction}
\ee


To put it in a nutshell, the multicenter bubbling solutions are characterized by the harmonic functions $\Gamma=(V,  K^1, \ldots , K^{n_v+1}; L_1 ,\ldots ,L_{n_v+1}, M)$ on $\mathbb{R}^3$. Schematically, we can write 
\be
\Gamma=\Gamma_\infty + \sum_{j=1}^n \frac{\Gamma_j}{\rho_j} \,.
\ee
One can define a symplectic product on $\mathbb{R}^{2n_V+4}$: for $A=(A^0,  A^1, \ldots , A^{n_v+1}; A_1 ,\ldots ,A_{n_v+1}, A_0)$ and $B=(B^0,  B^1, \ldots , B^{n_v+1}; B_1 ,\ldots ,B_{n_v+1}, B_0)$,
\be \langle A,B \rangle \equi A^0 B_0-A_0 B^0 + A^I B_I-A_I B^I \,. \ee

The absence of Dirac-Misner strings in the multicenter bubbling solutions then leads to conditions on the relative positions of the Gibbons-Hawking centres, the so-called \textit{bubble equations}, or \textit{Denef integrability equations} \cite{Denef:2000nb,Bena:2006is}:
\begin{equation}
\label{eq:BubbleEquations}
\sum_{j=1}^n \frac{ \langle \Gamma_i , \Gamma_j \rangle}{\rho_{ij}} \= \langle\Gamma_\infty , \Gamma_i \rangle \, , \qquad \textrm{for  } i\=1,\ldots n\,.
\end{equation}

\subsection{The STU model}
\label{subsubsec:the_STU_model}

The requirement that the five-dimensional geometry be asymptotically flat $\mathbb{R}^{1,4}$ constrains the asymptotic values of the harmonic functions $h_\infty$, $l_\infty$ and $k_\infty$ such that
\begin{equation}
V= \sum_{j=1}^n \frac{q_j}{\rho_j} \, , \qquad
L_I=1 +\sum_{j=1}^n \frac{l^I_j}{\rho_j} \, ,\qquad
K^I= \sum_{j=1}^n \frac{k^I_j}{\rho_j} \, , \qquad
M=m_\infty \+ \sum_{j=1}^n \frac{m_j}{\rho_j} \, .
\label{eq:harmfunc}
\end{equation}
Besides, requiring the resulting geometry to be smooth in five-dimensions amounts to constraining the values of the electric and momentum charges in terms of the magnetic and Kaluza-Klein monopole charges \cite{Bena:2005va,Berglund:2005vb}:
\be  \label{eq:smoothness_cond_GH}
l_j^I \=-\frac{1}{2}C_{IJK} \frac{k_j^J k_j^K}{q_j}\, ,\qquad m_j \= \frac{1}{12}C_{IJK} \frac{k_j^I k_j^J k_j^K}{q_j^2}\, .
\ee

These conditions allow each centre to preserve 16 supercharges; the overall solution, made of several centres, preserves 4 supercharges as the BMPV black hole \cite{Breckenridge:1996is}. It can be shown that this solution is equivalent to multiple stacks of D3-branes at angles in a T-dual frame \cite{Berglund:2005vb}.

The $\cN=2$ five-dimensional Supergravity coupled to $n_V=2$ extra vector fields has a metric and field strength (\ref{eq:c35dmetric&fields}) that simplify to three-charge solutions:
\be 
\begin{split}
ds_{5}^2 &\= -\left(Z_1 Z_2 Z_3\right)^{-\frac{2}{3}} \left(dt+\mu \left(d\psi + A \right) + \varpi \right)^2 \+ V^{-1} \left(Z_1 Z_2 Z_3\right)^{\frac{1}{3}} \, \left(d\psi + A\right)^2\\
& \hspace{0.55cm}\+ V \, \left(Z_1 Z_2 Z_3\right)^{\frac{1}{3}} \,\biggl[\,d\rho^2 + \rho^2 \, \left( d\vartheta^2 + \sin^2 \vartheta \, d\phi^2 \right)\,\biggr]\, ,\\
F^I &\= d_3 \left( Z_I^{-1} \left(dt+\omega \right) \right)\+ \Theta^I\,.
\label{eq:c3BubblingMetric}
\end{split}
\ee
This class of horizonless solutions have the same asymptotic geometry as the 4-supercharge five-dimensional rotating BMPV black holes \cite{Breckenridge:1996is}, which have a macroscopic horizon and are described by the harmonic functions
\be 
V \= \frac{1}{\rho}\, , \qquad L_I \= 1+\frac{Q_I}{\rho} \, , \qquad K^I \= 0 \,,\qquad M \= \frac{J_L}{\rho}\,.
\ee
Indeed, asympototically, these bubbling solutions behave like a BMPV black hole with charges $Q_I$, and left angular momentum $J_L$:
\be
\begin{split}
Q_I &\= \sum_{j=1}^n l_j^I \+ C_{IJK}\, \sum_{(i,j)=1}^n k_i^J k_j^K \, ,\\
J_L &\=  \frac{1}{2} \sum_{j=1}^n m_j \+ \frac{1}{2} \sum_{(i,j)=1}^n l_i^I k_j^I \+ \frac{C_{IJK}}{6}\sum_{(i,j,k)=1}^n k_i^I k_j^J k_k^K\,.
\label{eq:c3Asymptoticcharge5d}
\end{split}
\ee
In addition, the bubbling solutions have a right angular momentum $J_R$:
\be
j\equiv J_R \= \frac{1}{2} \left| \sum_{i<j}  \langle \Gamma_i , \Gamma_j \rangle \, \hat{\rho}_{ij}\right| \= \frac{1}{2} \left| \sum_{i}  \langle \Gamma_\infty , \Gamma_i \rangle \, \vec{\rho}_i \right| 
\,,\qquad \textrm{with  } \hat{\rho}_{ij} \:\equiv\:\frac{\vec{\rho}_i - \vec{\rho}_j}{| \vec{\rho}_i - \vec{\rho}_j|} \,.
\ee
Note that the left angular momentum $J_L$ is the one on the $\psi$-fiber, whereas the right angular momentum $J_R$ is understood as the angular momentum on $\mathbb{R}^3$. The BMPV black hole does not have any right angular momentum. Hence, as expected, by taking the scaling limit of multi-centered solutions, $J_R$ vanishes.

The \textit{scaling limit} is defined as the limit where the inter-centre distances $\rho_{ij}$ (between centres $i$ and $j$) shrinks uniformly to zero. The limit is parameterized by the \textit{scaling parameter}, $\lambda$: $\rho_{ij}=\lambda d_{ij}$, with $\max d_{ij}\equiv d = \cO(1)$.

Thus, given some charges $Q_I$ and an angular momentum $J_L$ for a BMPV black hole, there are various horizonless smooth bubbling solutions of $n$ Gibbons-Hawking centres that have the same asymptotic charges as the BMPV black hole. Counting how many of these solutions there are decomposes into two steps. The first step is to count the number of possibilities for the charges of the GH centres $\Gamma_j=(q_j,k^I_j;l^I_j,m_j)$ such that their asyptotic charges  $(Q_I,J_L)$ matches the BMPV black hole's. Then, each charge configuration $(\Gamma_j)_{j=1,\dots,n}$ defines a superselection sector\footnote{The term \textit{superselection sector} is here used in the sense that microstate geometries with different $\Gamma_j$ fluxes --- which are quantized --- cannot be related from one to another by moving in the moduli space of solutions \cite{Bena:2013gma}, except by quantum tunnelling.} that possesses a connected moduli space of solutions, whose quantization gives the number of states in that particular superselection sector.
Now, right angular-momentum of the bubbling solutions, $J_R$, is generically different from 0 (the value for the BMPV black hole). So, if one wishes to count the number of states in a superselection sector that have $J_R$ smaller than a threshold value $\varepsilon$, then one should apply the quantization procedure only in the region of moduli space where $J_R<\varepsilon$, which is in the vicinity of the scaling limit. 

\section{Kaluza-Klein modes at the scaling limit}
\label{sec:3}

The scaling limit is a point at the boundary of moduli space that plays a special role in the construction of microstate geometries. In order to understand the general shape of the vicinity of the scaling limit, there are two interesting questions. The first is whether the volume of the entire moduli space is finite. The second is whether the distance in moduli space between the scaling limit and any point in the bulk moduli space is finite or infinite. As distance on moduli space is a notion that arises in the Swampland context, we will try to tackle the second question through the lens of the Swampland programme.


\subsection{The length of the $\adstwo$ throat in terms of the scaling parameter}

It has been mentioned in the Introduction that in the scaling limit, the throat of microstate geometries deepens. 
We would like to estimate the length of the throat of a near-scaling solution presented in section \ref{subsubsec:the_STU_model}, in terms of the scaling parameter $\lambda$. We will compute the length of the throat of bubbling solutions approaching the scaling limit, and compare its behaviour with respect to $\lambda$ with the logarithmic divergence of the throat length of the BMPV black hole. The position of the centres in bubbling solutions is arbitrary (insofar as they satisfy the bubble equations), and different centre configurations will modify the throat length; however, we will show that this modification is set by the (coordinate) size of the region containing the centres.

For a BMPV black hole, the (radial) throat length is infinite, with the divergence being logarithmic. In other terms, let us fix a coordinate $\rho_M$ not too far at infinity ($\rho_M < Q_I$); its distance to a near-horizon cut-off $\rho_0$ is
\be \label{eq:throat_length_BMPV}
L_\text{throat}^\text{BMPV}(\rho_0,\rho_M) = \int_{\rho_0}^{\rho_M}V^{1/2}\left( Z_1Z_2Z_3 \right)^{1/6} \d \rho
\underset{\rho_0\rightarrow 0}{=} \left( Q_1Q_2Q_3 \right)^{1/6} \ln{\left(\frac{\rho_M}{\rho_0}\right)}
+ F(\rho_M)  \,.
\ee
The correction $F(\rho_M)$ is of order $\cO\left(\frac{\rho_M}{Q_I}\right)$ and induced by the constant term in $Z_I$. At first order in $\rho_M$, it is equal to $\frac{\rho_M-\rho_0}{Q_\textrm{har}}$, where $Q_\textrm{har}$ is the harmonic mean of $(Q_1,Q_2, Q_3)$. This correction behaves like a constant as $\rho_0$ approaches 0.
\newline

Now, consider a family of smooth multi-center bubbling solution in 5D approaching the scaling limit. The Gibbons-Hawking centres are at a coordinate distance $\rho_{ij}(\lambda)=\lambda d_{ij}$ of each other, where $d\equiv \max d_{ij}$ is of order 1. We choose the origin of the coordinates such that all the GH centres are within a radius of $\rho=\lambda d$. We want to know how the throat length scales with the scaling parameter $\lambda$. 
The throat length shall be computed from the same $\rho_M < Q_I$ to a region at the bottom of the throat at coordinate $\rho_0(\lambda)>\lambda d$. 
We shall define for instance
\be
\rho_0(\lambda)=2 \max_{i,j} \rho_{ij}(\lambda) =2 \lambda d \,,
\ee
so that we are looking at the distance between the asymptotics and the blob of GH centres. The point is to keep some distance with respect to each individual GH centres. 
The length of the throat in question is
\be \label{eq:throat_length_1}
L_\text{throat}(\rho_0(\lambda),\rho_M) = \int_{\rho_0(\lambda)}^{\rho_M}V^{1/2}\left( Z_1Z_2Z_3 \right)^{1/6} \d \rho \,.
\ee

As the scaling parameter $\lambda$ is sent to zero, the metric of a bubbling solution approaches that of a BMPV black hole. The more we are away from the bottom of the throat, the better the BMPV black hole approximation to the bubbling solution is.
%
More precisely, in the integration domain of the integral (\ref{eq:throat_length_1}), $\rho_j=\rho+\mathcal{O}(\lambda d)$, and $\frac{1}{\rho_j}=\frac{1}{\rho}\(1+\cO\(\frac{\lambda d}{\rho}\)\)$; so the function $Z_I V$ approximates to
\be
\label{eq:ZV_approx}
Z_IV=\frac{Q_I+\cO(\lambda d \times \mathrm{charges})}{\rho^2}+\frac{1}{\rho} \,.
\ee
The integrand of (\ref{eq:throat_length_1}) is then
\be \label{eq:sqrt_grr}
V^{1/2}\left( Z_1Z_2Z_3 \right)^{1/6}=
\frac{\left( Q_1Q_2Q_3 \right)^{1/6}}{\rho}
\left[ 1+\cO\left(\frac{\rho}{Q_I}\right) + \cO\left(\frac{\lambda d}{\textrm{charges}}\right) \right] \,.
\ee
The first correction to the logarithm comes from the asymptotic behaviour dominated by the $1/\rho$ term in (\ref{eq:ZV_approx}), and is exactly the same one as for the BMPV black hole. The second correction comes from the fact that the centres are arbitrarily distributed in a region of radius $\lambda d$.
Therefore, integrating the dominant term and its corrections leads to the following reorganization of terms \footnote{Note that we could have taken $\rho_M$ arbitrarily big. The important point is that integrating the $\cO\left(\frac{\rho}{Q_I}\right)$ term in (\ref{eq:sqrt_grr}) gives exactly the function $F(\rho_M)$ appearing in (\ref{eq:throat_length_BMPV}).}:
\be
\label{eq:L_throat_log_lambda}
L_\mathrm{throat}(\rho_0(\lambda),\rho_M) \underset{\lambda \rightarrow 0}{=}
-\left( Q_1Q_2Q_3 \right)^{1/6} \ln{\left(\frac{2 d \lambda}{\rho_M} \right)}
+ F(\rho_M)
+ \cO\(\frac{\lambda d}{\mathrm{charges}}\)\, \ln{\left(\frac{2 d \lambda}{\rho_M} \right)}
 \,.
\ee
Although the positions of the Gibbons-Hawking centres are arbitrary, they lie in a small region inside $\rho < \lambda d$. So in the scaling limit, they give rise to the geometry of the BMPV black hole outside of the blob region ($\rho\geq 2 \lambda d$), only up to small corrections. These are dominated by a $\lambda\ln(\lambda)$ term whose limit is zero. It was important to know that this correction's limit is zero, so that inverting equation (\ref{eq:L_throat_log_lambda}) gives
\be
\label{eq:rho0_VS_Lthroat}
\rho_0(\lambda) \underset{\lambda \rightarrow 0}{\sim} \rho_M \exp\( -\frac{L_\mathrm{throat}(\rho_0(\lambda),\rho_M) -F(\rho_M)}{(Q_1Q_2Q_3)^{1/6}} \) \,.
\ee

\subsection{The $\adstwo$ throat and Kaluza-Klein modes}


In this section we compute the mass scale of the $\sthree$ Kaluza-Klein towers. 
The five-dimensional metric (\ref{eq:c3BubblingMetric}) asymptotes to the $\adstwo \times \sthree$ metric in the throat region. 
Of course, near the Gibbons-Hawking centres, the geometry differs, but as long as we do not approach the GH centres too closely (for example $\rho\geq 10d\lambda$),
\be
Z_I = \frac{Q_I}{\rho} \left(1+\cO(\rho) \right) \,, \qquad 
V = \frac{1}{\rho} \left(1+\cO(\rho) \right) \,,
\ee
so that we get the metric of an $\adstwo \times \sthree$ up to
\be
\begin{split}
ds_{5}^2 = &-\[ \(Q_1 Q_2 Q_3\)^{-\frac{2}{3}}\rho^2 + \cO(\rho^3) \] \(dt+\omega \)^2 
+ \[\(Q_1 Q_2 Q_3\)^{\frac{1}{3}} \frac{1}{\rho^2} +\cO\(\frac{1}{\rho} \) \] d\rho^2 \\
&+ \[\(Q_1 Q_2 Q_3\)^{\frac{1}{3}} +\cO(\rho) \] \[\(d\psi + A\)^2 + d\vartheta^2 + \sin^2 \vartheta \, d\phi^2 \] \,.
\end{split}
\ee
At the location where $\rho=\rho_0(\lambda)$, which, can be understood being roughly the ``bottom of the throat'', there is an infinite tower of Kaluza-Klein modes on the $\sthree$, with the lightest mass measured at the bottom of the throat being
\be
m_{K K}(\rho_0(\lambda))\propto \frac{1}{R_{S^{3}}(\rho_0(\lambda))} \approx \frac{1}{\left(Q_{1} Q_{2} Q_{3}\right)^{\frac{1}{6}}}\,.
\ee
The mass of the $n^{\mathrm{th}}$ Kaluza-Klein mode measured at infinity gets redshifted to
\be
\label{eq:Mn_redshift_rho0}
M_n=n\, m_{K K} \sqrt{g_{tt}}|_{\rho=\rho_{0}(\lambda)}= n\, m_{K K}\left(Z_{1} Z_{2} Z_{3}\right)^{-1 / 3}|_{\rho=\rho_{0}(\lambda)}
\approx \frac{n \,\rho_{0}(\lambda)}{\left(Q_{1} Q_{2} Q_{3}\right)^{1 / 2}} \,.
\ee

Injecting (\ref{eq:rho0_VS_Lthroat}) into (\ref{eq:Mn_redshift_rho0}), we deduce that the tower of Kaluza-Klein states have masses that scale like
\be
\label{eq:masstower_Lthroat}
M_n(L_\text{throat}) \underset{\lambda \rightarrow 0}{\approx}
\frac{n \, G(\rho_M) }{(Q_1Q_2Q_3)^{1/2}} \exp\( -\frac{L_\text{throat}}{(Q_1Q_2Q_3)^{1/6}} \) \,,
\ee
where $G(\rho_M)=\rho_M \exp\(\frac{F(\rho_M)}{(Q_1Q_2Q_3)^{1/6}}\)$ \footnote{The approximation sign $\approx$ in (\ref{eq:masstower_Lthroat}) is here for the factor $\frac{n \, G(\rho_M) }{(Q_1Q_2Q_3)^{1/2}}$ in front of the exponential, but the exponential is exact.} .
This decreasing exponential mass is consistent with the extension of the Swampland Distance Conjecture to this system that we discussed in the Introduction.

\subsection{M2 branes at the bottom of the throat}
\label{subsection:3.3}

More generally, any locus in the cap verifies the approximation $\rho_i\ll Q_I$, so although $g_{tt}$ is not constant in the cap, its dependence with respect to the scaling parameter $\lambda$ is the same everywhere in the cap, leading to the same redshift behavior
\be
\sqrt{g_{tt}}|_{\rho_i=\lambda d_i} \underset{\lambda\rightarrow0}{\sim} f(d_i) \, \lambda \,.
\ee
The proportionality factor, $f(d_i)$, depends on the location of the point in the cap and is set by the charges $\Gamma_j$.

As a result, M2-branes wrapping the two-cycle linking two Gibbons-Hawking centres will experience a redshift that globally scales like $\lambda$ in the scaling limit, so using (\ref{eq:rho0_VS_Lthroat}) and dropping the proportionnality constant gives
\be
M_\mathrm{M2} \underset{\lambda\rightarrow0}{\sim} \exp\( -\frac{L_\text{throat}}{(Q_1Q_2Q_3)^{1/6}} \) \,.
\ee
We meet again the same exponential mass decrease for the tower of M2 branes.

As we go into the scaling limit and the throat becomes longer and longer, the M2 branes become also exponentially light.
\newline

Our system can be used to extend the Swampland Distance hypothesis. In our example, we move in the moduli space of metrics. In the scaling limit, the asymptotic geometry is unchanged. Besides, the size of the cap remains constant, as well as the inter-center physical proper distances \cite{Bena:2007qc}, up to order $\cO(\lambda)$: the geometry of the cap remains also fixed. In the scaling limit, the only modulus we are moving is the throat length which grows to infinity. 

Let $p_0$ a point in moduli space (a reference point), characterising a solution that possesses a throat region; and $\{p(\lambda)\}_{\lambda\in(0,\lambda_0]}$ the set of points in moduli space approaching the scaling limit ($\lambda\rightarrow0$) from $p(\lambda_0)=p_0$.
By reading off the argument inside the exponential, one possible conclusion is that the distance in moduli space between $p_0$ and $p(\lambda)$ should be proportional to the length of $p(\lambda)$'s throat, $L_\text{throat}(\lambda)$:
\be \label{eq:Delta_Swampland_throat_QQQ}
\alpha \, \Delta_\mathrm{exponential}(p_0,p(\lambda)) \underset{\lambda\rightarrow0}{\sim} 
\frac{L_\text{throat}(\rho_0(\lambda),\rho_M)}{(Q_1Q_2Q_3)^{1/6}} \,,
\ee
where $\alpha$ corresponds to the mass decay rate of the Swampland Distance Conjecture in (\ref{eq:Swampland_distance_conjecture}).
The distance to the scaling limit would then be \textit{infinite}. 

Note that, instead of having $n$ BPS Gibbons-Hawking centres coming closer to reach the scaling limit, the limit of $n$ coincident BPS black holes in $\cN=1$ Supergravity merging together lies also at \textit{infinite} distance in moduli space \cite{Michelson:1999dx}. The computation leading to (\ref{eq:masstower_Lthroat}) does not require the horizonless regularity conditions (\ref{eq:smoothness_cond_GH}) at each centre, and Gibbons-Hawking centres with a horizon going to the scaling limit are actually merging black holes. Our results thus agree with the infinite distance in moduli space in \cite{Michelson:1999dx}.
However, it is not clear that those two computations should give the same result. Indeed, the bubble equations (\ref{eq:BubbleEquations}) constrains the relative position of the ``Denef black holes'' (the Gibbons-Hawking centres with a horizon) from one another to be dependent of the charges $\Gamma_i$; whereas there is no such a constraint on the relative position of the ``Michelson-Strominger black holes'' of \cite{Michelson:1999dx}.
\newline

Interestingly, $(Q_1Q_2Q_3)^{1/6}$ is approximately the radius of the 3-sphere in the regime $\rho_0(\lambda)< \rho < Q_I$. 
Indeed, the radius of the 3-sphere is $(Q_1Q_2Q_3)^{1/6}$ up to corrections of order $\cO\(\frac{\rho}{Q_I}\)$ near $\rho \sim Q_I$, and corrections of order the magnitude of the charges $\Gamma_j$ in the vicinity of $\rho\sim \rho_0(\lambda)$; so the throat looks very much like a cylinder with an $\sthree$ base. Let us define the aspect ratio $\mathcal{R}$ of the throat to be the throat length divided by the radius of the $\sthree$ base.
Then 
\be \label{eq:Delta_Swampland_throat}
\alpha \, \Delta_\mathrm{exponential}(p_0,p(\lambda)) \underset{\lambda\rightarrow0}{\sim} 
\frac{L_\text{throat}(\rho_0(\lambda),\rho_M)}{R_{S^{3}}}
= \mathcal{R}(\lambda) \,.
\ee
Note that reading off the argument of the decreasing exponential gives only the distance in moduli space in the vicinity of the scaling limit, and only in the direction towards the scaling limit; we do not have any piece of information about how the distance behaves near $p_0$.

\section{Distance on the Phase space of Multi-centered bubbling solutions}
\label{sec:4}

The dimensional reduction of the smooth five-dimensional Supergravity solutions of Sections \ref{sec:2} and \ref{sec:3} along the $\psi$-fiber leads to the four-dimensional multi-centered solutions \cite{Bates:2003vx}. Describing these centres at equilibrium separations from each other (\ref{eq:BubbleEquations}) from Supergravity at $g_s N\gg 1$ is related to the quiver description of wrapped D-branes at $g_s N\ll 1$ \cite{Denef:2002ru}. As mentioned in the Introduction, one can compute the symplectic form from the quiver description, and, when possible, use the compatible complex structure to define a distance on moduli space, $\Delta_\mathrm{phase}$.

In this section, we wish to check whether the distance $\Delta_\mathrm{phase}$ coincides with the distance obtained by reading off the exponential decrease.


%
%
%

\subsection{Symplectic form from Quiver Quantum Mechanics}
\label{subsec:SymplecticFormQuiverQM}

Given $L$, the Lagrangian governing the dynamics of $n$-centered bubbling solutions of four-dimensional $\cN=2$ Supergravity, coupled to $n_V$ gauge fields,
and given $\phi^A$ a basis of the fields apprearing in the Lagrangian, the symplectic form of the Supergravity-solution space is defined by
\be \label{eq:symplectic_form_1}
\Omega \equiv \int d\Sigma_l \,\, \delta\left( \frac{\partial L}{\partial (\partial_l \phi^A)}\right) \wedge \delta \phi^A\, ,
\ee
where $\Sigma$ is a Cauchy surface (in the 4-dimensional spacetime).
We consider $\tilde{\Omega}$ the restriction of the symplectic form $\Omega$ to the space of multicentered solutions (which verify the bubble equations). This consists of changing and restricting the variable fields $\phi^A$, such that the new fields $\phi'^I$ define the $2n-2$-dimensional configuration of the $n$ GH centres.

The symplectic form of BPS solutions in Supergravity is difficult to compute for multi-centre solutions through the Supergravity action. Nevertheless, in \cite{deBoer:2008zn}, the authors computed the symplectic form in the open string description, valid when the centres do not backreact ($g_sN\ll1$). Thanks to a non-renormalization theorem in a similar spirit as \cite{Denef:2002ru}, this symplectic form is independent of $g_s$ and equal to the symplectic form of BPS Supergravity solutions.

Indeed, the authors of \cite{deBoer:2008zn,Denef:2002ru,Denef:2000nb} argue that the open string dual of $n$ GH centers in Supergravity is described by supersymmetric vacua of a (0+1)-dimensional quiver gauge theory, whose Coulomb branch consists --- after integrating out the massive bifundamentals --- of $n$ abelian vector multiplets. Each of the vector multiplets comprises three scalars $(x^1,x^2,x^3)$ which characterize the positions of the D6 branes in $\mathbb{R}^3$, one auxiliary field, $D$, and one gauge field, $A$, which corresponds to the spatial components of the 4D gauge field $\mathcal{A}$ in Supergravity. The effective action of the vector multiplets in the Coulomb branch is determined by the Lagrangian
\be \label{effact}
L_{\textrm{quiver}}=\sum_{p=1}^{n} (-U_p D_p + A_p \cdot \dot{x}_p) + \,\,{\rm
fermions}\,\,+\,\, \textrm{higher-order terms},
\ee
where $U_p$ is found to be 
\be
U_p= \langle \Gamma_p,H_p(x_p) \rangle \equiv \left\langle \Gamma_p \,,\, \theta+\sum_{q\neq p} \frac{\Gamma_q}{|x_p-x_q|} \right\rangle \,.
\ee
The symplectic form can be extracted from $L_{\textrm{quiver}}$.

Applying (\ref{eq:symplectic_form_1}) to $L_{\textrm{quiver}}$, the authors of \cite{deBoer:2008zn} obtain the symplectic form to be of the form $\sum_p \delta x_p \wedge \delta A_p$. The restriction to BPS solutions corresponds, in the open string language, to restricting the solution space to $\bigcap_p \{ U_p=0\}$. In terms of the Supergravity data, the restricted symplectic form becomes
\be \label{symplform2}
\tilde{\Omega} = \frac{1}{2} \sum_p \delta x^i_p \wedge \langle \Gamma_p ,\delta \mathcal{A}_d^i(x_p)\rangle \,.
\ee
After calculations detailed in \cite{deBoer:2008zn}, the infinitesimal variations of the field $\delta \mathcal{A}_d^i(x_p)$ in (\ref{symplform2}) can be replaced by infinitesimal variations of the locations of the GH centres $\delta {\mathbf x}_p$, such that
\be \label{con3}
\tilde{\Omega} = \frac{1}{4} \sum_{p\neq q} \langle \Gamma_p,\Gamma_q\rangle
\frac{\epsilon_{ijk} (\delta (x_p-x_q)^i \wedge \delta(x_p-x_q)^j) \, (x_p-x_q)^k }{|{\mathbf x}_p - {\mathbf x}_q|^3} \,.
\ee
Because the GH centres satisfy the bubble equations (\ref{eq:BubbleEquations}), acting on the positions ${\mathbf x}_p$ of a solution with $\mathrm{SO}(3)$ rotations gives another configuration satisfying the bubble equations. Thus, if we impose the variations of the positions of the GH centres to be an infinitesimal rotation along the $\mathbf{n}$-axis as $\delta x_p^i = \epsilon^{iab} n^a x_p^b$, and call $X_n$ the vector field corresponding to the rotation, then the reduced symplectic form satisfies
\be \label{sympl_form_under_rotation}
\tilde{\Omega}(X_n,\cdot)  =  n^i \delta J^i \,,
\ee
where $J^i$ are the components of the angular momentum vector
\be \label{eq:J_components}
J^i = \frac{1}{4} \sum_{p\neq q} \langle \Gamma_p,\Gamma_q\rangle \frac{x_p^i-x_q^i }{|{\mathbf x}_p - {\mathbf x}_q|} \,.
\ee
Using equation (\ref{sympl_form_under_rotation}), it is possible to deduce the whole reduced symplectic form for two and three GH centres.
Furthermore, the reduced symplectic form (\ref{con3}) is closed, so the $(2n-2)$-dimensional solution space can be viewed as a phase space.

\subsection{The moduli space of three-centre solutions}
\label{ssection:4.2}

We now specialize in a solution with three Gibbons-Hawking centres. 
In this superselection sector, the moduli space of solutions, which is also the phase space, has $2n-2=4$ dimensions. Here, we have already set the centre of mass of the three GH points to be at the origin of $\mathbb{R}^3$. The total angular momentum vector $J$ of the three-centre system (\ref{eq:J_components}) is described by its norm, $j$, and its direction --- parameterized by the $(\theta,\phi)$ angles in $\stwo$. Now, rotating the triangle formed by the GH centres around the axis of $J$ does not modify the angular momentum vector, so the fourth real variable that we call $\sigma$ characterizes this $\mathrm{U}(1)$ rotational symmetry.

In a nutshell, $(j,\theta,\phi,\sigma)$ are the coordinates on the four-dimensional phase space. Once the charges on each GH centre are fixed, the intersection products $\langle h, \Gamma_p \rangle$ and $\langle \Gamma_p, \Gamma_q \rangle$ are also fixed. Given the length of two sides of the triangle of the GH centres, the third one is determined by the bubble equations (\ref{eq:BubbleEquations}). In other terms, for a given size of the triangle, its shape is determined. And what controls the size of triangle in these coordinates is the angular momentum $j$ through 
\begin{equation}
j=\frac{1}{2}\sqrt{-\sum_{p<q}\langle h,\Gamma_p\rangle\langle h,\Gamma_q\rangle\,\rho_{pq}^2}\,,
\end{equation}
where $p$ and $q$ label the centres.
The angles $(\theta,\phi,\sigma)$ then parameterize how the triangle orients itself in $\mathbb{R}^3$. They do not change the nature of the bubbling solution, but they do contribute to the phase space of solutions. 
\newline

Using (\ref{sympl_form_under_rotation}), the symplectic form reduces to \cite{deBoer:2008zn}
\be \label{fsym}
\tilde{\Omega} = - \d (j\cos\theta)\wedge \d\phi - \d j \wedge \d\sigma \;.
\ee
We can then define $x\equiv j$ and $y\equiv j\cos{\theta}$, so that 
\be \label{sympl_form}
\tilde{\Omega} = - \d x \wedge \d\sigma - \d y \wedge \d\phi \;.
\ee
Note that $(x,y;\sigma,\theta)$ are the symplectic (or action-angle) coordinates, since in these coordinates, the symplectic form $\tilde{\Omega}$ is flat.

\begin{figure}[h]
\begin{center}
\includegraphics[page=5,scale=1]{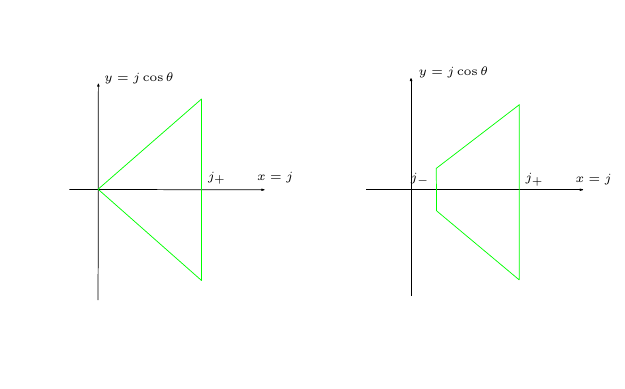}
\caption{The moduli space on the $(x,y)$-plane is a polytope, deliminated by the facets (they are edges too) of the polytope. The picture on the left depicts the moduli space when $j_- = 0$ while the one on the right applies when $j_- > 0$. We only consider the former instance. The scaling limit lies at point $(0,0)$. The figures are from \cite{deBoer:2008zn}. }\label{fig_polytope}
\end{center}
\end{figure}

On the one hand, the ``action'' coordinates $x$ and $y$ satisfy the following inequalities
\be 
x-j_-\geq 0, \quad j_+-x\geq 0, \quad x-y\geq 0,\quad x+y\geq 0. \label{polytope}
\ee
So on the $(x,y)$-plane, the solution space is a convex polytope. In particular, when $j_-=0$, this is a triangle, as shown in figure \ref{fig_polytope}.
On the other hand, the ``angle'' coordinates define a two-torus fibration over the polytope. On each facet of the polytope, one of the angles $\sigma$ or $\phi$ becomes degenerate, that is to say the cycle they parameterize shrinks to zero size. Indeed, on the $x-y = 0$ and $x+y=0$ facets, $\phi$ becomes degenerate, whereas on the $x-j_-= 0$ and $j_+-x= 0$ facets, $\sigma$ becomes degenerate. 

We consider the instance $j_-=0$ where scaling solutions are admitted. The polytope of the $(x,y)$ moduli space is defined by 3 inequalities defining a triangle:
\be
l_1(x,y)= -x+j_+ \geq 0 \qquad l_2(x,y)=x-y \geq 0 \qquad l_3(x,y)=x+y \geq 0 \,.
\ee
The scaling limit lies at $(0,0)$ in the $(x,y)$-plane. The scaling limit loci lie at $(0,0,\sigma,\phi)$.
As both $\phi$ and $\sigma$ become degenerate at the scaling point $(0,0)$, this shows the scaling limit boils down to one point, at the boundary of moduli space.

The phase space is a convex toric manifold endowed with a closed symplectic form, 
and the vertices lie on integer coordinates
(so our polytope is a 2-dimensional Delzant polytope~\footnote{Further details on Kähler toric manifolds and definition of Delzant polytopes can be found in Appendix B of \cite{deBoer:2008zn}}).
Consequently, the symplectic form is compatible with an integrable complex structure, so that our phase space is a K\"ahler manifold \cite{2000math......4122A}. Thus, we can use the following result for Kähler toric manifolds to determine the K\"ahler metric:

With the set of $p$ inequalities characterizing the (two-dimensional) polytope $P$ of the Kähler toric manifold, 
\be
    l_i(\textbf{x}) = 
     c^x_{i}x + c^y_{i}y - \lambda_i \geq 0 \,, 
\label{fun}
\ee
one can define the ``canonical potential'' of the polytope $P$
\be \label{eq:canonical_potential_toric_manifold}
   g_P(\textbf{x}) = \frac{1}{2} \sum_{r=1}^{p} l_r (\textbf{x}) \log l_r(\textbf{x})\,
\ee
whose Hessian $G=\left( \frac{\partial^2g_P}{\partial x_i \partial x_j} \right)_{i,j}$ in the ``action'' coordinates determines the complex K\"ahler structure in the action-angle coordinates \cite{2000math......4122A,guillemin1994}
\be
J=
\begin{pmatrix}
   0 & \rvline & -G^{-1}\\
   \hline
G & \rvline & 0
 \end{pmatrix} \,.
\label{complx_struct_general}
\ee
Thus the Riemannian Kähler metric in the action-angle coordinates is
\be
\tilde{\Omega} (\cdot,J\cdot)=
G_{ij} \d x^i\otimes \d x^j + (G^{-1})_{ij} \d \theta^i \otimes \d \theta^j
=
\begin{pmatrix}
   G & \rvline & 0\\
   \hline
0 & \rvline & G^{-1}
 \end{pmatrix} \,.
\label{metric_general}
\ee

Consequently, we apply this result and deduce that the moduli space metric in the symplectic coordinates is of the form (\ref{metric_general}) with
\be
G
= \frac{1}{x^2-y^2}
\begin{pmatrix}
   \frac{2xj_+-x^2-y^2}{2(j_+-x)} & -y\\
    -y & x
 \end{pmatrix}
 \quad , \quad
 G^{-1}
 = \frac{1}{2j_+-x}
 \begin{pmatrix}
   2x(j_+-x) & 2y(j_+-x)\\
    2y(j_+-x) & 2xj_+-x^2-y^2
 \end{pmatrix}\,.
\ee

We immediatly see that the $G$ part of the Riemannian metric blows up on all the facets of the triangle. Thus, the metric is defined only in the interior of the triangle. In this regard, the scaling limit is not part of the bulk moduli space, but in its boundary.

Actually, the symplectic form on the K\"ahler toric manifold does not define a unique ``potential'' determining $J$. In fact, $J$ can be defined by any potential $g$ of the form
\be \label{eq:potential_g_general} g=g_P+h \,, \ee
where $h$ is a smooth function on the whole polytope $P$ satisfying the requirements that \cite{2000math......4122A}:
\begin{itemize}
    \item[\textbf{(1)}] the Hessian $G$ of $g$ is positive definite on the interior $P^\circ$ of $P$, and
    \item[\textbf{(2)}] the determinant of $G$ is of the form
\be
\det(G)=\gamma(\textbf{x})\( \prod_{r=1}^p l_r(\textbf{x}) \)^{-1} \,,
\ee
with $\gamma$ being a smooth and strictly positive function on the \textit{whole} $P$.
\end{itemize}
The Hessian $G=\left( \frac{\partial^2g}{\partial x_i \partial x_j} \right)_{i,j}$ defines the compatible toric complex structure $J$ and Riemannian K\"ahler metric $\tilde{\Omega} (\cdot,J\cdot)$ the same way as in using (\ref{complx_struct_general}) and (\ref{metric_general}).

Nevertheless, we will continue our computations with the metric defined by the ``canonical potential'' $g_P$, as did \cite{deBoer:2008zn}.

\subsection{The distance to the scaling limit}
\label{ssection:the_scaling_limit_in_moduli_space}

The volume of the moduli space of the bosonic sector of BPS, classical configurations of Supergravity (which is our phase space) naively counts, in units of the Planck constant $h^{n-1}$ (where $n$ is the number of GH centres), the number of quantum states in a particular superselection sector. Indeed,
\be
\mathcal{V}_{\mathrm{phase}}= \hbar^2 \int \d x \, \d y \, \d \sigma \, \d \phi \, \sqrt{\det(G  G^{-1)}} = h^2 {j_+}^2
\ee
in the instance where $j_-=0$. When $j_- \neq 0$, the number of states is ${j_+}^2-{j_-}^2$.
This naive counting, which does not include the fermionic degrees of freedom, matches nevertheless with the result in \cite{deBoer:2008zn}.
From the symplectic-form derivation, the volume of the entire moduli space is finite.

Imposing that the volume of the entire moduli space is finite has a consequence on the shape of the vicinity of scaling limit. 
Indeed, if the length to the scaling limit was infinite, then the area of its orthogonal directions should shrink at a rate such that the volume remains finite. Then, for a parametrically small angular momentum, $j$, in the classical regime, the density of quantum states at that given phase-space hypersurface (defining a given throat length) would be parametrically small. Each superselection sector's vicinity to the scaling limit would have the shape of a spike of infinite-length and finite volume.
%
\newline

Now we wish to assess whether the geodesic distance in solution space between the scaling limit and any point in the bulk moduli space is infinite.
The symplectic coordinates of the solution space are bounded, and the metric is not singular in its bulk, so any two points in the bulk solution space are at finite distance of each other. The only place where the distance could be infinite is at the facets of the polytope.

The coordinate values of $(\sigma,\phi)$ are chosen in $[0,2\pi]$ and their metric is bounded by $2j_+$ from above. We will therefore only consider the metric from the $(x,y)$ coordinates.

Consider the straight path between the scaling limit $\overrightarrow0=(0,0)$ and the point $\overrightarrow{r_0}=(x_0,y_0)=r_0(\cos\alpha,\sin\alpha)$, with $\alpha\in [-\pi/4,\pi/4]$. The distance of this path is given by
\be
\Delta_\mathrm{phase}(\overrightarrow0,\overrightarrow{r_0})=\int_{0}^{r_0}\sqrt{G_{ab}\frac{\d x^a}{\d r}\frac{\d x^b}{\d r}}\d r \,.
\ee
In the vicinity of the scaling point, although the metric blows up ($G_{ab}\frac{\d x^a}{\d r}\frac{\d x^b}{\d r} \underset{r\rightarrow 0}{\sim} \frac{\cos\alpha}{r}$), to compute the distance we integrate its square root:
\be
\label{eq:scaling_point_distance}
\Delta_\mathrm{phase}(\overrightarrow0,\overrightarrow{r_0}) \underset{r_0\rightarrow 0}{\sim} 2\sqrt{\cos\alpha} \, \sqrt{r_0} = 2\sqrt{x_0} \;.
\ee
Then the distance to the scaling limit $\Delta_\mathrm{phase}(\overrightarrow0,\overrightarrow{r_0})$ is \textit{finite}; therefore the geodesic distance in moduli space is finite too. This contradicts the naive extension of the Swampland Distance Conjecture.

With similar reasoning, we can show that although the metric is blowing up on the facets of polytope in the $(x,y)$-plane, the entire moduli space is bounded. The details are in the Appendix.
\newline


We wish now to relate the distance on moduli space with the masses $\sthree$ Kaluza-Klein modes.
Recall that the angular momentum $j$ is proportional to the scaling parameter $\lambda$:
\begin{equation}
j=\frac{1}{2}\sqrt{-\sum_{a<b}\langle h,\Gamma_a\rangle\langle h,\Gamma_b\rangle\,\rho_{ab}(\lambda)^2}\,,\label{angularnorm}
\end{equation}
where $\rho_{ab}(\lambda)=\lambda d_{ab}$ is the coordinate distance between centres $a$ and $b$.
We can therefore express the masses of the Kaluza-Klein modes (\ref{eq:Mn_redshift_rho0}) in terms of the angular momentum $j$:
\be
M_n(j)=\kappa n j + \cO(j^2) \;,
\ee
where $\kappa$ is a positive constant
\be
\frac{1}{\kappa} \approx \frac{1}{4d}\left(Q_{1} Q_{2} Q_{3}\right)^{1 / 2} \sqrt{-\sum_{a<b}\langle h,\Gamma_a\rangle\langle h,\Gamma_b\rangle\,d_{ab}^2} \,.
\ee
As we have seen in equation (\ref{eq:scaling_point_distance}), the distance in moduli space from any point $p$ at finite angular momentum $j$ to the scaling limit $j = 0$ is finite. We deduce that as $j$ approaches 0, the mass of the Kaluza-Klein tower depends quadratically on this distance:
\be
   M_n(j) \underset{j\rightarrow 0}{\sim} \frac{\kappa n}{4} \Delta_\mathrm{phase}(0,j)^2 \;.
\ee
This quadratic dependence deviates from the exponential dependence advocated by the Swampland Distance Conjecture.\footnote{Since the distance to the scaling limit is finite here, we are measuring distances from the moving point $p(j)$ to the scaling limit $p(0)$; while in Section \ref{sec:3}, as the distance to the scaling limit was infinite, we were considering distances from the moving point $p(j)$ to a reference point $p(j_0)$ at finite angular momentum.} This indicates that the distance of the moduli space seen as a phase space computed using the non-renormalization theorem and the canonical potential --- according to \cite{deBoer:2008zn} --- differs from the distance one should use in order to extend the Swampland Distance Conjecture.

\section{Discussion}
\label{sec:5}

In this paper, we have focused our attention on the scaling limit of a class of microstate geometries --- the bubbling solutions --- in the moduli space of solutions. This limit plays a particular role, as microstates geometries approach the black hole solution from the asymptotic observer's perspective.
As one moves towards the scaling limit, bubbling solutions develop a throat whose depth is increasing to infinity. Besides, the deepening of the throat makes the redshift from the cap to the spatial asymptotics stronger and stronger, so the energy of all excitations lying at the bottom of the throat decrease to zero. 

This decrease of energy excitations at the bottom of the throat is independent of the type of excitation we consider, as the redshift affecting them, set by $\sqrt{g_{tt}}$, is the same. In Section \ref{sec:3}, we have proved that the redshift decreases the energy excitations by a factor of $\exp\( -\frac{L_\text{throat}}{(Q_1Q_2Q_3)^{1/6}} \)$. Thus, one may argue that our model is a new instance of the Swampland Distance Conjecture for metrics. If it turns out to be true, one can extract, from the mass decay, a notion of distance in moduli space, $\Delta_\mathrm{exponential}\(p(\lambda_0), p(\lambda)\)$, from a reference solution $p(\lambda_0)$ to a solution $p(\lambda)$ approaching the scaling limit ($\lambda \rightarrow 0$).

As discussed in Section \ref{subsection:3.3}, the Michelson-Strominger derivation of the distance in moduli space \cite{Michelson:1999dx} shows that the merging of $n$ BPS black holes happens as well at an \textit{infinite} moduli space distance from the bulk, and seems to support the $\Delta_\mathrm{exponential}$ distance. However, it is not clear that these two distances --- one involving black holes with unconstrained positions, and the other involving charged Gibbons-Hawking centres/black holes whose positions satisfy the Denef integrability equations  --- should agree.

A second notion of distance, $\Delta_\mathrm{phase}$, can be derived from the K\"ahler metric of the phase space of three-centre solutions. This distance is \textit{a priori} computed in the weak coupling regime. The first question is whether one can extrapolate this distance up to strong string coupling regime. The non-renormalization theorem of \cite{deBoer:2008zn} shows that the reduced symplectic form, $\tilde{\Omega}$, (\ref{symplform2}) remains the same in the Supergravity regime up to a normalization factor. Nevertheless, the potential $g_P$ (\ref{eq:canonical_potential_toric_manifold}) used to compute the integrable structure $J$ is not unique (\ref{eq:potential_g_general}) --- and is so in \textit{all} regimes of the string coupling. As a result, in order to assert that the complex structure, $J$, and the metric of the moduli space are invariant under the tuning of the string coupling, one must show that the effects of $h$ in (\ref{eq:potential_g_general}) on the metric on moduli space are dominated by those of the canonical potential, $g_P$.

We have shown that there exists a tension between the ``canonical'' distance according to the phase-space computation, $\Delta_\mathrm{phase}$, whose distance to the scaling limit is finite, and $\Delta_\mathrm{exponential}$. Now, there is only one correct normalization of the distance on the moduli space of bubbling solutions at strong string coupling: the one from the variations of the effective Supergravity action. Thus, we have the following possibilities:
%
\begin{itemize}
    \item[\textbf{(1)}] Neither $\Delta_\mathrm{phase}$ nor $\Delta_\mathrm{exponential}$ give the correct normalization.
    \item[\textbf{(2)}] Only the canonical $\Delta_\mathrm{phase}$ gives the correct distance on moduli space, even in the strong string coupling regime. If the Swampland Distance Hypothesis for metrics is correct, then it will not apply to our metrics.
   \item[\textbf{(3)}] The Swampland Distance Hypothesis applies to our solutions, and $\Delta_\mathrm{exponential}$ gives the correct normalization of the distance to the scaling limit. The use of canonical $\Delta_\mathrm{phase}$ is not reliable in the strong coupling regime.
\end{itemize}

\vspace{11pt}


If possibility (2) is correct, then our computation gives an explicit example of a metric on moduli space which blows up at all points on the boundary of moduli space, but where all of the boundary points lie at finite distance in moduli space. In particular, the scaling limit of bubbling solutions --- at which global symmetries of the Black hole are restored --- is within \textit{finite}-distance reach from any other point in the moduli space. Besides, the mass decay of the tower of Kaluza-Klein modes does not behave like a decreasing exponential with respect to the moduli space distance $\Delta_\mathrm{phase}$ between $p(\lambda_0)$ and $p(\lambda)$.

Although there is an infinite tower of states whose mass is decaying to zero, the decay is due to an universal redshift in a fixed-warp region of space-time, and thus does not introduce any singularities.\footnote{Actually, this argument does not depend on assuming possibility (2).} Besides, the three-sphere at the bottom of the throat on which the Kaluza-Klein modes live is macroscopic and is part and parcel of the five-dimensional Supergravity solution. Therefore, the example we provide here differs from the usual Swampland picture, in which going at a corner in moduli space implies the appearance of singularities (for instance the shrinking of Calabi-Yau cycles in \cite{Lee:2018urn}), which involve the breakdown of the effective field theory.

As a result, the Swampland conjectures would not forbid the scaling limit to be accessible from the bulk moduli space. 
However, in possibility (2), as argued in \cite{deBoer:2008zn}, quantum mechanics, by virtue of the uncertainty principle, will imply the breakdown of Supergravity at the scaling limit. 
\newline

If possibility (3) is correct, then one cannot extend the ``canonical'' $\Delta_\mathrm{phase}$ to the strong coupling regime, because the integrable complex structure, $J$, is not invariant under the shift of $g_s$, or because one has to take into account the effect of the additional potential $h$ at weak coupling in the first place. However, the authors of \cite{deBoer:2008zn} computed $J$ in the open string picture with the canonical potential $g_P$, and used it at strong coupling regime. In particular, the probability distribution $\mathrm{e}^{-\mathcal{K}}$ of quantum wave functions in the phase space of bubbling solution they derive depends on the K\"ahler potential $\mathcal{K}$, whose value will shift if one considers the potential $h$ in addition of the canonical potential $g_P$.

Therefore, if the canonical $\Delta_\mathrm{phase}$ somehow gives the wrong normalization of distance, some of the conclusions in \cite{deBoer:2008zn} could be revisited. 
The relative coordinate positions of the centers $\overrightarrow{\rho_{ij}}$ define solutions in Supergravity. In particular, near the scaling limit, the coordinate positions $\overrightarrow{\rho_{ij}}$ need to be arbitrarily precise. However, because of the form of the symplectic form (\ref{con3}) computed from the open string sector, these coordinates do not commute; hence, it is not possible to localize the positions $\overrightarrow{\rho_{ij}}$ with arbitrarily good precision in \textit{coordinate space}. In the closed string sector, the fully back-reacted solution does not require its Gibbons-Hawking centres to be localized with arbitrarily good precision in the geometry in terms of \textit{proper distance}, as the cap keeps its shape in the scaling limit. However, if one follows the logic of \cite{deBoer:2008zn}, the uncertainty about positions in coordinate space at weak string coupling is transported unto the phase space of three-centered solutions in Supergravity: one cannot localize any classical Supergravity bubbling solution with arbitrary high precision in phase space. Instead, each classical Supergravity bubbling solution is defined with some inherent quantum uncertainty and must be coarse-grained with a ``droplet'' of solutions around it in a volume $h^{m}$ in the phase space. How far in the phase space one should apply the coarse-graining depends on the metric/distance in moduli space around that particular classical Supergravity solution $p_1$. 

On the one hand, when the components of the moduli-space metric have small values, as one schematically moves away from $p_1$ within the coarse-graining droplet region of $p_1$, one can reach solutions that are very different from $p_1$. In particular, according to the \textit{canonical} distance on moduli space that \cite{deBoer:2008zn} used --- where the metric behaves like $1/J_R$ in the vicinity of the scaling limit --- the scaling limit lies at \textit{finite} distance to any other point in the moduli space, so the coarse-graining of a solution $p_1$ close to the scaling limit point contains solutions $\{p\}$ which possess throats that have very different macroscopic physical lengths. Therefore, Heisenberg's uncertainty principle prevents classical solutions from Supergravity to be a good description of black hole microstates.

On the other hand, when the components of the moduli-space metric are large around the solution $p_1$, the solutions $\{p\}$ reached within a distance $\sim \sqrt{h}$ look much more like $p_1$. In particular, with a metric that, in the vicinity of the scaling limit, scales like $1/{{J_R}^2}$ \footnote{This gives the logarithmic dependence of the distance on $\lambda$ in (\ref{eq:L_throat_log_lambda})} as advocated by the Swampland Distance Hypothesis, the distance in moduli space to the scaling limit is \textit{infinite}. Wandering around $p_1$ within a distance $\sqrt{h}$ along the angular-momentum coordinate $J_R$ (or equivalently the scaling-parameter coordinate $\lambda$) cannot give solutions $\{p\}$ whose physical throat length is arbitrarily long.
Instead, with $\Delta_\mathrm{exponential}$, one deduces from (\ref{eq:Delta_Swampland_throat}) that the variation of the length of the throat in the set of solutions $\{p\}$ will be of order
\be
\Delta L_\mathrm{throat}= \alpha \, R_{S^{3}} \,.
\ee
Whether quantum fluctuations in $p_1$'s coarse-graining droplet are negligible or too large depends on the value of the mass decay rate $\alpha$ of (\ref{eq:Delta_Swampland_throat}). If $\alpha \ll 1$, the geometries described by Supergravity are reliable and well-defined. If $\alpha$ is of order one or bigger however --- as in the context of Calabi-Yau compactifications \cite{Klaewer:2016kiy,Gendler:2020dfp,Andriot:2020lea} ---, quantum fluctuations of the throat length of each bubbling solution have macroscopic size, so describing those arbitrarily deep geometries with Supergravity is still not reliable. Nonetheless, in both instances, the breakdown (should it happen) of Supergravity here is milder than the one from the canonical phase space distance of \cite{deBoer:2008zn}.

In a nutshell, regardless of the value of $\alpha$, coarse-graining droplets defined using $\Delta_\mathrm{exponential}$ contain a much smaller range of solutions than those using the canonical $\Delta_\mathrm{phase}$. 
While with the canonical $\Delta_\mathrm{phase}$ as the correct normalization of distance on the moduli space, the coarse-graining droplet of a deep-throat bubbling solution could contain the scaling limit point; the droplet derived from the $\Delta_\mathrm{exponential}$ normalization only contains solutions with similar throat lengths. Therefore, with $\Delta_\mathrm{exponential}$, the breakdown of Supergravity at the scaling limit is softened.
\newline

If extending the Swampland Distance Hypothesis to our model is possible, then we have drawn a parallel between (i) travelling within Planckian field range in field space to avoid the breakdown of the EFT in the context of the Swampland Distance Hypothesis and (ii) travelling a distance of $\sqrt{h}$ around a classical solution in phase space within the region of its quantum fluctuations. While the Swampland Distance Hypothesis only considers field ranges that are isotropic in moduli space, the fundamental quantity on the phase space side is the coarse-grained volume $h^m$, and the symplectic form defines an anisotropic ``droplet'' around a classical solution. 
In our example, the Swampland Distance Hypothesis could be interpreted as a consequence of the symplectic form establishing Heisenberg's uncertainty principle. Whether or not this interpretation is legitimate is a question to explore.
\newline

As for Black Hole physics, both possibilities (2) and (3) entail at least \textit{some} breakdown of Supergravity as the description of arbitrarily-deep-throat bubbling solutions. As one approaches the scaling limit from a bubbling solution, the resulting geometry enters into a new phase, whose precise description may require other tools, for instance perturbative String Theory --- that one uses in the microstate solutions of \cite{Martinec:2015pfa,Martinec:2017ztd,Martinec:2018nco,Martinec:2019wzw,Martinec:2020gkv}.
However, the extent of the breakdown, depending on which one of possibility (2) or (3) is correct, is very different. 

With possibility (2), the bubbling solution acquires a critical maximal throat length after which supergravity completely breaks down. Thus, the new phase can possess a throat that is not arbitrarily deep, like in the instances of \cite{Giusto:2004id,Giusto:2004ip,Giusto:2004kj,Giusto:2012yz} and \cite{Martinec:2017ztd,Martinec:2018nco,Martinec:2019wzw,Martinec:2020gkv}.

With possibility (3), when the Supergravity description of a geometry with a very long throat (of length $L_\mathrm{throat}$) becomes unreliable, one has to scramble the initial Supergravity solution with solutions whose throat length is between $L_\mathrm{throat}-\alpha \, R_{S^{3}}$ and $L_\mathrm{throat}+\alpha \, R_{S^{3}}$. Therefore, the new phase should still possess a throat which can be tuned to be arbitrarily deep. Thus, any complete description of bubbling solutions up to the scaling limit should still capture the presence of a cap and an arbitrarily deep throat.
Finding such a description beyond Supergravity of those geometries would then be an interesting direction for the future.





\acknowledgments
I would like to thank Iosif Bena, Guillaume Bossard, Mariana Gra\~na, \'Alvaro Herr\'aez, Elias Kiritsis, Dieter L\"ust, Severin L\"ust, Daniel Mayerson, Ruben Monten, Cumrun Vafa and Nick Warner for useful discussions and references.
This work was partially supported
by the ERC Consolidator Grant 772408-Stringlandscape, the ERC Advanced Grant 787320 - QBH Structure, the ANR grant Black-dS-String ANR-16-CE31-0004-01 and the John Templeton Foundation grant 61169.

\appendix

\section{Appendix: Boundedness of the moduli space of 3-centre solutions from the phase space distance}

To show that the entire moduli space is bounded using the canonical $\Delta_\mathrm{phase}$, we will probe the asymptotic behaviour of the metric at all the different facets and vertices of the polytope: the vertex at $(0,0)$, $(j_+,j_+)$ and $(j_+,-j_+)$, and the facets at $x=j_+$, $x-y=0$ and $x+y=0$.
\newline

\textbf{The vertex at $(0,0)$ (the scaling limit).} This instance has already been studied in the subsection \ref{ssection:4.2}.
\newline

\textbf{The facets $x-y=0$ and $x+y=0$.} 
Given $x_F\in (0,j_+)$, we approach any point $M_F(x_F,\pm x_F)$ on the facets by a straight horizontal line from a point $M_0(x_0,\pm x_F)$ in the bulk. On the path, at the point $M(x,\pm x_F)$,
\be
G_{xx}\underset{M \rightarrow M_F}{\sim} \frac{\jp \xe - \xe^2}{2(j_+-\xe)\xe} \, \frac{1}{x-x_F}= \frac{1}{2(x-x_F)}\;
\ee
giving a square-root behaviour to the path distance
\be
\Delta(M_F,M_0) \underset{M_0 \rightarrow M_F}{\sim} \, \sqrt{2(x_0-x_F)} \;.
\ee
Therefore the distance in moduli space is finite.

Note that this computation does not take into account the scaling-limit point, as we used $\xe\neq 0$ in our equations.
\newline

\textbf{The facet $x=j_+$.} Given $y_F\in (-\jp,j_+)$, we approach any point $M_F(\jp,y_F)$ on the facet by a straight horizontal line from a point $M_0(x_0, y_F)$ in the bulk. On the path, at the point $M(x,y_F)$,
\be
G_{xx}\underset{M \rightarrow M_F}{\sim} \frac{1}{2(j_+-x)} \;
\ee
gives a square root behaviour to the path distance
\be
\Delta(M_F,M_0) \underset{M_0 \rightarrow M_F}{\sim}  \sqrt{2(j_+-x)} \;.
\ee
Therefore the geodesic distance in moduli space is finite.
\newline

\textbf{The vertex at $(j_+,j_+)$.} 
We approach this limit from the point $M_0$ of coordinates $\left(\jp-r_0\cos\alpha,\jp-r_0\sin\alpha\right)$, with $\alpha \in (\pi/4,\pi/2)$. Near the vertex, the metric behaves like
\be
G_{ij}\frac{\d x^i}{\d r}\frac{\d x^j}{\d r} \underset{r\rightarrow 0}{\sim} \frac{\sin\alpha}{2r} \;,
\ee
so the path length is finite.
\newline

\textbf{The vertex at $(j_+,-j_+)$.}
We approach this limit from the point $M_0$ of coordinates $\left(\jp-r_0\cos\alpha,-\jp+r_0\sin\alpha\right)$, with $\alpha \in (\pi/4,\pi/2)$. Near the vertex, the metric also behaves like
\be
G_{ij}\frac{\d x^i}{\d r}\frac{\d x^j}{\d r} \underset{r\rightarrow 0}{\sim} \frac{\sin\alpha}{2r} \;,
\ee
so the path length is finite.

\newpage

\bibliographystyle{utphys}      
\bibliography{Deep_throat_revelations}

\providecommand{\href}[2]{#2}\begingroup\raggedright\begin{thebibliography}{10}

\bibitem{Mathur:2005zp}
S.~D. Mathur, ``{The Fuzzball proposal for black holes: An Elementary
  review},'' \href{http://dx.doi.org/10.1002/prop.200410203}{{\em Fortsch.
  Phys.} {\bfseries 53} (2005) 793--827},
  \href{http://arxiv.org/abs/hep-th/0502050}{{\ttfamily arXiv:hep-th/0502050}}.

\bibitem{Skenderis:2008qn}
K.~Skenderis and M.~Taylor, ``{The fuzzball proposal for black holes},''
  \href{http://dx.doi.org/10.1016/j.physrep.2008.08.001}{{\em Phys. Rept.}
  {\bfseries 467} (2008) 117--171},
  \href{http://arxiv.org/abs/0804.0552}{{\ttfamily arXiv:0804.0552 [hep-th]}}.

\bibitem{Avery:2009tu}
S.~G. Avery, B.~D. Chowdhury, and S.~D. Mathur, ``{Emission from the D1D5
  CFT},'' \href{http://dx.doi.org/10.1088/1126-6708/2009/10/065}{{\em JHEP}
  {\bfseries 10} (2009) 065}, \href{http://arxiv.org/abs/0906.2015}{{\ttfamily
  arXiv:0906.2015 [hep-th]}}.

\bibitem{Bena:2007kg}
I.~Bena and N.~P. Warner, ``{Black holes, black rings and their microstates},''
  \href{http://dx.doi.org/10.1007/978-3-540-79523-0\_1}{{\em Lect. Notes Phys.}
  {\bfseries 755} (2008) 1--92},
  \href{http://arxiv.org/abs/hep-th/0701216}{{\ttfamily arXiv:hep-th/0701216}}.

\bibitem{Warner:2019jll}
N.~P. Warner, ``{Lectures on Microstate Geometries},''
  \href{http://arxiv.org/abs/1912.13108}{{\ttfamily arXiv:1912.13108
  [hep-th]}}.

\bibitem{Heidmann:2019gvg}
P.~Heidmann, {\em {Black-Hole Microstates in String Theory: Black is the Color
  but Smooth are the Geometries?}}
\newblock PhD thesis, Orsay, 2019.

\bibitem{Bena:2006kb}
I.~Bena, C.-W. Wang, and N.~P. Warner, ``{Mergers and typical black hole
  microstates},'' \href{http://dx.doi.org/10.1088/1126-6708/2006/11/042}{{\em
  JHEP} {\bfseries 11} (2006) 042},
  \href{http://arxiv.org/abs/hep-th/0608217}{{\ttfamily arXiv:hep-th/0608217}}.

\bibitem{Bena:2015bea}
I.~Bena, S.~Giusto, R.~Russo, M.~Shigemori, and N.~P. Warner, ``{Habemus
  Superstratum! A constructive proof of the existence of superstrata},''
  \href{http://dx.doi.org/10.1007/JHEP05(2015)110}{{\em JHEP} {\bfseries 05}
  (2015) 110}, \href{http://arxiv.org/abs/1503.01463}{{\ttfamily
  arXiv:1503.01463 [hep-th]}}.

\bibitem{Bena:2016ypk}
I.~Bena, S.~Giusto, E.~J. Martinec, R.~Russo, M.~Shigemori, D.~Turton, and
  N.~P. Warner, ``{Smooth horizonless geometries deep inside the black-hole
  regime},'' \href{http://dx.doi.org/10.1103/PhysRevLett.117.201601}{{\em Phys.
  Rev. Lett.} {\bfseries 117} no.~20, (2016) 201601},
  \href{http://arxiv.org/abs/1607.03908}{{\ttfamily arXiv:1607.03908
  [hep-th]}}.

\bibitem{Bena:2013gma}
I.~Bena, A.~Puhm, O.~Vasilakis, and N.~P. Warner, ``{Almost BPS but still not
  renormalized},'' \href{http://dx.doi.org/10.1007/JHEP09(2013)062}{{\em JHEP}
  {\bfseries 09} (2013) 062}, \href{http://arxiv.org/abs/1303.0841}{{\ttfamily
  arXiv:1303.0841 [hep-th]}}.

\bibitem{Bena:2009ev}
I.~Bena, G.~Dall'Agata, S.~Giusto, C.~Ruef, and N.~P. Warner, ``{Non-BPS Black
  Rings and Black Holes in Taub-NUT},''
  \href{http://dx.doi.org/10.1088/1126-6708/2009/06/015}{{\em JHEP} {\bfseries
  06} (2009) 015}, \href{http://arxiv.org/abs/0902.4526}{{\ttfamily
  arXiv:0902.4526 [hep-th]}}.

\bibitem{Bena:2009en}
I.~Bena, S.~Giusto, C.~Ruef, and N.~P. Warner, ``{Multi-Center non-BPS Black
  Holes: the Solution},''
  \href{http://dx.doi.org/10.1088/1126-6708/2009/11/032}{{\em JHEP} {\bfseries
  11} (2009) 032}, \href{http://arxiv.org/abs/0908.2121}{{\ttfamily
  arXiv:0908.2121 [hep-th]}}.

\bibitem{Bah:2021jno}
I.~Bah, I.~Bena, P.~Heidmann, Y.~Li, and D.~R. Mayerson, ``{Gravitational
  Footprints of Black Holes and Their Microstate Geometries},''
  \href{http://arxiv.org/abs/2104.10686}{{\ttfamily arXiv:2104.10686
  [hep-th]}}.

\bibitem{Bena:2018bbd}
I.~Bena, P.~Heidmann, and D.~Turton, ``{AdS$_{2}$ holography: mind the cap},''
  \href{http://dx.doi.org/10.1007/JHEP12(2018)028}{{\em JHEP} {\bfseries 12}
  (2018) 028}, \href{http://arxiv.org/abs/1806.02834}{{\ttfamily
  arXiv:1806.02834 [hep-th]}}.

\bibitem{Ooguri:2006in}
H.~Ooguri and C.~Vafa, ``{On the Geometry of the String Landscape and the
  Swampland},'' \href{http://dx.doi.org/10.1016/j.nuclphysb.2006.10.033}{{\em
  Nucl. Phys. B} {\bfseries 766} (2007) 21--33},
  \href{http://arxiv.org/abs/hep-th/0605264}{{\ttfamily arXiv:hep-th/0605264}}.

\bibitem{Grimm:2018ohb}
T.~W. Grimm, E.~Palti, and I.~Valenzuela, ``{Infinite Distances in Field Space
  and Massless Towers of States},''
  \href{http://dx.doi.org/10.1007/JHEP08(2018)143}{{\em JHEP} {\bfseries 08}
  (2018) 143}, \href{http://arxiv.org/abs/1802.08264}{{\ttfamily
  arXiv:1802.08264 [hep-th]}}.

\bibitem{Lust:2019zwm}
D.~Lüst, E.~Palti, and C.~Vafa, ``{AdS and the Swampland},''
  \href{http://dx.doi.org/10.1016/j.physletb.2019.134867}{{\em Phys. Lett. B}
  {\bfseries 797} (2019) 134867},
  \href{http://arxiv.org/abs/1906.05225}{{\ttfamily arXiv:1906.05225
  [hep-th]}}.

\bibitem{DeWitt:1967yk}
B.~S. DeWitt, ``{Quantum Theory of Gravity. 1. The Canonical Theory},''
  \href{http://dx.doi.org/10.1103/PhysRev.160.1113}{{\em Phys. Rev.} {\bfseries
  160} (1967) 1113--1148}.

\bibitem{Candelas:1990pi}
P.~Candelas and X.~de~la Ossa, ``{Moduli Space of {Calabi-Yau} Manifolds},''
  \href{http://dx.doi.org/10.1016/0550-3213(91)90122-E}{{\em Nucl. Phys. B}
  {\bfseries 355} (1991) 455--481}.

\bibitem{Bonnefoy:2019nzv}
Q.~Bonnefoy, L.~Ciambelli, D.~Lüst, and S.~Lüst, ``{Infinite Black Hole
  Entropies at Infinite Distances and Tower of States},''
\href{http://arxiv.org/abs/1912.07453}{{\ttfamily arXiv:1912.07453 [hep-th]}}.

\bibitem{Denef:2000nb}
F.~Denef, ``{Supergravity flows and D-brane stability},''
  \href{http://dx.doi.org/10.1088/1126-6708/2000/08/050}{{\em JHEP} {\bfseries
  08} (2000) 050}, \href{http://arxiv.org/abs/hep-th/0005049}{{\ttfamily
  arXiv:hep-th/0005049}}.

\bibitem{Bates:2003vx}
B.~Bates and F.~Denef, ``{Exact solutions for supersymmetric stationary black
  hole composites},'' \href{http://dx.doi.org/10.1007/JHEP11(2011)127}{{\em
  JHEP} {\bfseries 11} (2011) 127},
  \href{http://arxiv.org/abs/hep-th/0304094}{{\ttfamily arXiv:hep-th/0304094}}.

\bibitem{Breckenridge:1996is}
J.~Breckenridge, R.~C. Myers, A.~Peet, and C.~Vafa, ``{D-branes and spinning
  black holes},'' \href{http://dx.doi.org/10.1016/S0370-2693(96)01460-8}{{\em
  Phys. Lett. B} {\bfseries 391} (1997) 93--98},
  \href{http://arxiv.org/abs/hep-th/9602065}{{\ttfamily arXiv:hep-th/9602065}}.

\bibitem{Witten:1986qs}
E.~Witten, ``{Interacting Field Theory of Open Superstrings},''
  \href{http://dx.doi.org/10.1016/0550-3213(86)90298-1}{{\em Nucl. Phys. B}
  {\bfseries 276} (1986) 291--324}.

\bibitem{Crnkovic:1987tz}
C.~Crnkovic, ``{Symplectic Geometry of the Covariant Phase Space, Superstrings
  and Superspace},'' \href{http://dx.doi.org/10.1088/0264-9381/5/12/008}{{\em
  Class. Quant. Grav.} {\bfseries 5} (1988) 1557--1575}.

\bibitem{deBoer:2008zn}
J.~de~Boer, S.~El-Showk, I.~Messamah, and D.~Van~den Bleeken, ``Quantizing n=2
  multicenter solutions,''
  \href{http://dx.doi.org/10.1088/1126-6708/2009/05/002}{{\em JHEP} {\bfseries
  05} (2009) 002}, \href{http://arxiv.org/abs/0807.4556}{{\ttfamily
  arXiv:0807.4556 [hep-th]}}.

\bibitem{Gutowski:2004yv}
J.~B. Gutowski and H.~S. Reall, ``{General supersymmetric AdS(5) black
  holes},'' \href{http://dx.doi.org/10.1088/1126-6708/2004/04/048}{{\em JHEP}
  {\bfseries 04} (2004) 048},
  \href{http://arxiv.org/abs/hep-th/0401129}{{\ttfamily arXiv:hep-th/0401129}}.

\bibitem{Bena:2004de}
I.~Bena and N.~P. Warner, ``{One ring to rule them all ... and in the darkness
  bind them?},'' \href{http://dx.doi.org/10.4310/ATMP.2005.v9.n5.a1}{{\em Adv.
  Theor. Math. Phys.} {\bfseries 9} no.~5, (2005) 667--701},
  \href{http://arxiv.org/abs/hep-th/0408106}{{\ttfamily arXiv:hep-th/0408106}}.

\bibitem{Bena:2006is}
I.~Bena, C.-W. Wang, and N.~P. Warner, ``{The Foaming three-charge black
  hole},'' \href{http://dx.doi.org/10.1103/PhysRevD.75.124026}{{\em Phys. Rev.
  D} {\bfseries 75} (2007) 124026},
  \href{http://arxiv.org/abs/hep-th/0604110}{{\ttfamily arXiv:hep-th/0604110}}.

\bibitem{Bena:2005va}
I.~Bena and N.~P. Warner, ``{Bubbling supertubes and foaming black holes},''
  \href{http://dx.doi.org/10.1103/PhysRevD.74.066001}{{\em Phys. Rev. D}
  {\bfseries 74} (2006) 066001},
  \href{http://arxiv.org/abs/hep-th/0505166}{{\ttfamily arXiv:hep-th/0505166}}.

\bibitem{Berglund:2005vb}
P.~Berglund, E.~G. Gimon, and T.~S. Levi, ``{Supergravity microstates for BPS
  black holes and black rings},''
  \href{http://dx.doi.org/10.1088/1126-6708/2006/06/007}{{\em JHEP} {\bfseries
  06} (2006) 007}, \href{http://arxiv.org/abs/hep-th/0505167}{{\ttfamily
  arXiv:hep-th/0505167}}.

\bibitem{Bena:2007qc}
I.~Bena, C.-W. Wang, and N.~P. Warner, ``{Plumbing the Abyss: Black ring
  microstates},'' \href{http://dx.doi.org/10.1088/1126-6708/2008/07/019}{{\em
  JHEP} {\bfseries 07} (2008) 019},
  \href{http://arxiv.org/abs/0706.3786}{{\ttfamily arXiv:0706.3786 [hep-th]}}.

\bibitem{Michelson:1999dx}
J.~Michelson and A.~Strominger, ``{Superconformal multiblack hole quantum
  mechanics},'' \href{http://dx.doi.org/10.1088/1126-6708/1999/09/005}{{\em
  JHEP} {\bfseries 09} (1999) 005},
\href{http://arxiv.org/abs/hep-th/9908044}{{\ttfamily arXiv:hep-th/9908044
  [hep-th]}}.

\bibitem{Denef:2002ru}
F.~Denef, ``Quantum quivers and hall / hole halos,''
  \href{http://dx.doi.org/10.1088/1126-6708/2002/10/023}{{\em JHEP} {\bfseries
  10} (2002) 023}, \href{http://arxiv.org/abs/hep-th/0206072}{{\ttfamily
  arXiv:hep-th/0206072}}.

\bibitem{2000math......4122A}
M.~{Abreu}, ``{Kahler geometry of toric manifolds in symplectic coordinates},''
  {\em arXiv Mathematics e-prints} (Apr, 2000) math/0004122,
  \href{http://arxiv.org/abs/math/0004122}{{\ttfamily arXiv:math/0004122
  [math.DG]}}.

\bibitem{guillemin1994}
V.~Guillemin, ``Kaehler structures on toric varieties,''
  \href{http://dx.doi.org/10.4310/jdg/1214455538}{{\em J. Differential Geom.}
  {\bfseries 40} no.~2, (1994) 285--309}.
  \url{https://doi.org/10.4310/jdg/1214455538}.

\bibitem{Lee:2018urn}
S.-J. Lee, W.~Lerche, and T.~Weigand, ``{Tensionless Strings and the Weak
  Gravity Conjecture},'' \href{http://dx.doi.org/10.1007/JHEP10(2018)164}{{\em
  JHEP} {\bfseries 10} (2018) 164},
  \href{http://arxiv.org/abs/1808.05958}{{\ttfamily arXiv:1808.05958
  [hep-th]}}.

\bibitem{Klaewer:2016kiy}
D.~Klaewer and E.~Palti, ``{Super-Planckian Spatial Field Variations and
  Quantum Gravity},'' \href{http://dx.doi.org/10.1007/JHEP01(2017)088}{{\em
  JHEP} {\bfseries 01} (2017) 088},
  \href{http://arxiv.org/abs/1610.00010}{{\ttfamily arXiv:1610.00010
  [hep-th]}}.

\bibitem{Gendler:2020dfp}
N.~Gendler and I.~Valenzuela, ``{Merging the Weak Gravity and Distance
  Conjectures Using BPS Extremal Black Holes},''
  \href{http://arxiv.org/abs/2004.10768}{{\ttfamily arXiv:2004.10768
  [hep-th]}}.

\bibitem{Andriot:2020lea}
D.~Andriot, N.~Cribiori, and D.~Erkinger, ``{The web of swampland conjectures
  and the TCC bound},'' \href{http://dx.doi.org/10.1007/JHEP07(2020)162}{{\em
  JHEP} {\bfseries 07} (2020) 162},
  \href{http://arxiv.org/abs/2004.00030}{{\ttfamily arXiv:2004.00030
  [hep-th]}}.

\bibitem{Martinec:2015pfa}
E.~J. Martinec and B.~E. Niehoff, ``{Hair-brane Ideas on the Horizon},''
  \href{http://dx.doi.org/10.1007/JHEP11(2015)195}{{\em JHEP} {\bfseries 11}
  (2015) 195}, \href{http://arxiv.org/abs/1509.00044}{{\ttfamily
  arXiv:1509.00044 [hep-th]}}.

\bibitem{Martinec:2017ztd}
E.~J. Martinec and S.~Massai, ``{String Theory of Supertubes},''
  \href{http://dx.doi.org/10.1007/JHEP07(2018)163}{{\em JHEP} {\bfseries 07}
  (2018) 163}, \href{http://arxiv.org/abs/1705.10844}{{\ttfamily
  arXiv:1705.10844 [hep-th]}}.

\bibitem{Martinec:2018nco}
E.~J. Martinec, S.~Massai, and D.~Turton, ``{String dynamics in NS5-F1-P
  geometries},'' \href{http://dx.doi.org/10.1007/JHEP09(2018)031}{{\em JHEP}
  {\bfseries 09} (2018) 031}, \href{http://arxiv.org/abs/1803.08505}{{\ttfamily
  arXiv:1803.08505 [hep-th]}}.

\bibitem{Martinec:2019wzw}
E.~J. Martinec, S.~Massai, and D.~Turton, ``{Little Strings, Long Strings, and
  Fuzzballs},'' \href{http://dx.doi.org/10.1007/JHEP11(2019)019}{{\em JHEP}
  {\bfseries 11} (2019) 019}, \href{http://arxiv.org/abs/1906.11473}{{\ttfamily
  arXiv:1906.11473 [hep-th]}}.

\bibitem{Martinec:2020gkv}
E.~J. Martinec, S.~Massai, and D.~Turton, ``{Stringy Structure at the BPS
  Bound},'' \href{http://arxiv.org/abs/2005.12344}{{\ttfamily arXiv:2005.12344
  [hep-th]}}.

\bibitem{Giusto:2004id}
S.~Giusto, S.~D. Mathur, and A.~Saxena, ``{Dual geometries for a set of
  3-charge microstates},''
  \href{http://dx.doi.org/10.1016/j.nuclphysb.2004.09.001}{{\em Nucl. Phys. B}
  {\bfseries 701} (2004) 357--379},
  \href{http://arxiv.org/abs/hep-th/0405017}{{\ttfamily arXiv:hep-th/0405017}}.

\bibitem{Giusto:2004ip}
S.~Giusto, S.~D. Mathur, and A.~Saxena, ``{3-charge geometries and their CFT
  duals},'' \href{http://dx.doi.org/10.1016/j.nuclphysb.2005.01.009}{{\em Nucl.
  Phys. B} {\bfseries 710} (2005) 425--463},
  \href{http://arxiv.org/abs/hep-th/0406103}{{\ttfamily arXiv:hep-th/0406103}}.

\bibitem{Giusto:2004kj}
S.~Giusto and S.~D. Mathur, ``{Geometry of D1-D5-P bound states},''
  \href{http://dx.doi.org/10.1016/j.nuclphysb.2005.09.037}{{\em Nucl. Phys. B}
  {\bfseries 729} (2005) 203--220},
  \href{http://arxiv.org/abs/hep-th/0409067}{{\ttfamily arXiv:hep-th/0409067}}.

\bibitem{Giusto:2012yz}
S.~Giusto, O.~Lunin, S.~D. Mathur, and D.~Turton, ``{D1-D5-P microstates at the
  cap},'' \href{http://dx.doi.org/10.1007/JHEP02(2013)050}{{\em JHEP}
  {\bfseries 02} (2013) 050}, \href{http://arxiv.org/abs/1211.0306}{{\ttfamily
  arXiv:1211.0306 [hep-th]}}.

\end{thebibliography}\endgroup

\end{document}